\documentclass[12pt,a4paper, leqno]{article}
\usepackage[utf8]{inputenc}
\usepackage[T1]{fontenc}
\usepackage{float}
\usepackage{graphicx,float,afterpage,lscape, adjustbox}
\usepackage{lscape, rotating}
\usepackage{caption}
\usepackage[table,dvipsnames]{xcolor}
\usepackage[ruled,vlined]{algorithm2e}
\usepackage{endnotes}
\usepackage{tikz}
\RequirePackage{pifont,latexsym,ifthen,theorem,rotating,calc,textcase,booktabs}
\RequirePackage{amsfonts,amssymb,amsbsy,amsmath}
\RequirePackage[errorshow]{tracefnt}
\RequirePackage[onehalfspacing]{setspace}
\usepackage{caption,subcaption}
 \usepackage{booktabs}
\usepackage{graphicx}
\usepackage{caption}

\newtheorem{proposition}{Proposition}
\newtheorem{remark}{Remark}

\newtheorem{assumption}{Assumption}

\usepackage[title]{}
\RequirePackage[colorlinks=true,urlcolor=blue, citecolor=blue, linkcolor= black,hidelinks,pdftex]{hyperref}
\hypersetup{colorlinks=true,urlcolor=blue, citecolor=blue, linkcolor= blue}
\usepackage{cleveref}

	\usetikzlibrary{matrix,chains,positioning,decorations.pathreplacing,arrows}
\usetikzlibrary{positioning,calc}
\usepackage{tikz}
\usetikzlibrary{chains, positioning, fit, arrows.meta, shapes,calc}

\RequirePackage{xcolor}
\RequirePackage{bbm}
\RequirePackage{natbib} %% uncomment this for author-year bibliography

\usetikzlibrary{calc}
\usetikzlibrary{arrows}
\definecolor{NL}{HTML}{0072BD}
\definecolor{L}{HTML}{D95319}
\definecolor{S}{HTML}{EDB120}

\tikzset{%
	line numbers/.store in=\fakelinenos,
	line numbers=50,
	line number shift/.store in=\fakelinenoshift,
	line number shift=5mm,
	line number style/.style={text=gray},
}

\usepackage{doi-frbc}
\begin{document}

\title{Deep Neural Network Estimation in Panel Data Models\thanks{The views expressed herein are those of the authors and not necessarily those of the Federal Reserve Bank of Cleveland or the Federal Reserve System.}}
\author{{ Ilias Chronopoulos\thanks{{\small Essex Business School, University of Essex, Email: ilias.chronopoulos@essex.ac.uk}}}
\and { Katerina Chrysikou\thanks{{\small King's Business School, King's
College London, Email: katerina.chrysikou@kcl.ac.uk}}}
\and { George Kapetanios \thanks{{\small King's Business School, King's
College London, Email: george.kapetanios@kcl.ac.uk}}}
\and { James Mitchell \thanks{{\small Federal Reserve Bank of Cleveland, Email: james.mitchell@clev.frb.org}}}
\and { Aristeidis Raftapostolos \thanks{{\small King's Business School, King's
College London, Email: aristeidis.1.raftapostolos@kcl.ac.uk}}}}
\date{{ \today}}
\maketitle

\begin{abstract}
In this paper we study neural networks and  their approximating power in panel data models.  We  provide asymptotic guarantees on deep feed-forward neural network estimation of the conditional mean, building on the work of \cite{farrell2021deep}, and explore latent patterns in the cross-section.  We use the proposed estimators to forecast the progression of new COVID-19 cases across the G7 countries during the pandemic. We find significant forecasting gains over both linear panel and nonlinear time series models. Containment or lockdown policies, as instigated at the national-level by governments, are found to have out-of-sample predictive power for new COVID-19 cases.  We illustrate how the use of partial derivatives can help open the ``black-box'' of neural networks and facilitate semi-structural analysis: school and workplace closures are found to have been effective policies at restricting the progression of the pandemic across the G7 countries. But our methods illustrate significant heterogeneity and time-variation in the effectiveness of specific containment policies.

\[
\]
{\small \newline JEL codes: C33, C45. 
	  }
{\small \newline}

\end{abstract}

\newpage

			\section{Introduction}

			Panel  data models are widely used in economics and finance.  They combine both cross-sectional and time series data.  One important advantage of panel data over time-series methods (see, for example, Chapters 26 and 28 of \cite{pesaran2015time})    is their ability to control for unobserved heterogeneity both in the temporal and longitudinal dimensions. One can then approximate this latent individual heterogeneity through identifiable effects that are otherwise non-detectable in traditional time-series data sets.  There are several ways to model and control for individual heterogeneity in linear panel data models: the random effects estimator, see, for example,  \cite{balestra1966pooling},  the fixed effects (within) estimator, see, for example, \cite{mundlak1961empirical, mundlak1978pooling}, and the \cite{swamy1970efficient} estimator.  Alternative ways to model individual heterogeneity in linear models are found in  \cite{hsiao1974statistical, hsiao1975some}, with a thorough discussion in \cite{hsiao2004random, hsiao2008random} and Part VI of \cite{pesaran2015time}.
		
		The work summarized above focuses on linear heterogeneous panel data models.  However, the importance of nonlinearity has attracted increased interest in the literature.  Notable contributions are \cite{fernandez2016individual}, who adapt the analytical and jackknife bias correction methods introduced in \cite{hahn2004jackknife} to nonlinear models with additive or interactive individual and
		time effects, and \cite{CHEN2021296}  who address estimation and inference in general nonlinear models using iterative estimation.  \cite{haciouglu2021common}  provide an approach for estimation and inference in nonlinear
		conditional mean panel data models in the presence of cross-sectional dependence.     \cite{jochmans2017two}   develops the asymptotic properties of GMM estimators for models with two-way multiplicative fixed effects, while  \cite{charbonneau2013multiple} considers a logit conditional maximum likelihood approach to investigate whether existing panel methods for eliminating a single fixed effect can be modified to eliminate multiple fixed effects. 
		
		In this paper we also focus on the estimation of nonlinear panels. We propose the use of a novel machine learning (ML) panel data estimator based on neural networks. To help delineate the contributions of this paper and the empirical application that we consider, we first provide a high-level summary of the current literature on ML. 
		
		Statistical ML is  a major interdisciplinary research area. In the last decade, ML methods have been incorporated, in various forms, across the natural, social, medical, and economic sciences, leading to significant research outputs. There are two main reasons for such widespread adoption. Firstly, ML methods and specifically neural networks, the focus of this paper, have been found to exhibit outstanding empirical performance when forecasting, specifically with high-dimensional data sets.  Secondly, they have great capacity to uncover potentially unknown and both highly complicated and nonlinear relationships in the data. In conjunction with increased availability of high-dimensional data sets, and policymakers' understandable desire for accurate forecasts,  considerable attention has been paid to ML. 			
		
		Studies have shown that feed-forward neural networks
		can approximate any continuous function of several real variables arbitrarily well; see, for example, 
		\cite{hornik1991approximation}, \cite{hornik1989multilayer}, \cite{galant1992learning}, and \cite{par1991}.  Other
		nonparametric approaches, for example, splines, wavelets, the Fourier basis, as well as simple
		polynomial approximations, have the universal approximation property, based on the
		Stone--Weierstrass theorem. However, it has been convincingly argued that neural networks outperform them in prediction (see, for example, \cite{bla2010}).
		
		More recent work by \cite{liang2016deep} and \cite{YAROTSKY2017103, yarotsky2018optimal} considers feed-forward neural networks as approximations for complex functions that accommodate multiple layers, provided sufficiently
		many hidden neurons and layers are available. 
		Other examples, like   \cite{bartlett2019nearly},  provide the theoretical framework for neural network estimation, while \cite{Hieber2019} focuses on the adaptation property of neural networks, showing that they can strictly improve on classical methods. If the unknown target function is  a composition of simpler functions, then the composition-based deep net estimator is superior to estimators that do not use compositions.   Lastly, recent work of \cite{farrell2021deep}, building on the work of \cite{YAROTSKY2017103} and \cite{bartlett2019nearly}, studies deep neural networks and considers their use for semi-parametric inference. 
		\newline\indent
		In this paper we focus on nonlinear panel data models, where the source of nonlinearity lies in the conditional mean.   Our contribution  to the literature is as follows. We  propose a ML estimator of  the conditional mean,  $E(y_{it}|\boldsymbol{x}_{it})$,  based on neural networks and  explore the idea of heterogeneity in a nonlinear panel model by allowing the conditional mean to have a panel -- common nonlinear component -- as well as a nonlinear idiosyncratic component. We base our theoretical results mainly on  \cite{farrell2021deep}, expanding their contribution to a panel data framework. We also find evidence of the double descent effect, whereby complex models can perform well without the need for explicit regularization (see \cite{hastie2022surprises} and \cite{kelly2022virtue}, as well as Remark \ref{remark_double_descen} below).  

  We use the new deep panel data models to forecast the transmission of new COVID-19 cases during the pandemic across a number of countries. We consider the G7 countries. In contrast to theoretical epidemiological models, that may be specified incorrectly, our proposed neural network models are flexible reduced-form models. They let the data determine the path of new infections over time, by modeling this path as dependent on the lagged levels of the number of infections. By comparing the models against a deep (nonlinear) time-series model, that does not aim to exploit cross-country dependencies, we test whether there are benefits  when forecasting new COVID-19 cases to pooling data across countries. We find that there clearly are. Importantly, our model also captures the nonlinear features of a pandemic, particularly in its early waves. 

Neural networks have great capacity to approximate complicated nonlinear functions and have been found to forecast well. But they are frequently criticized as non-interpretable (of being a ``black box''), since they do not offer simple summaries of relationships in the data. Recently, there have been a number of papers that try to make ML output interpretable; see, for example, \cite{athey2017state}, \cite{wager2018estimation}, \cite{belloni2014inference},  \cite{joseph2019shapley}, \cite{chronopoulos2023forecasting}, and
\cite{kap2023}.

In this paper, given the many but contrasting (across time and countries) containment or social-distancing policies instigated to moderate the path of the COVID-19 pandemic, we use our model to shed light on the relative effectiveness -- across time and across the G7 countries -- of these policies at lowering the number of new COVID-19 cases. We do so by exploring how the use of partial derivatives, calculated from the output of our proposed neural network, can help examine the effectiveness of policy.  We examine the derivatives over time and find that some, but not all, containment policies were effective at lowering new COVID-19 cases. These policies tended to be more effective two to three weeks after the policy change.  There is also considerable heterogeneity across countries in the effectiveness of these policies. Policy, as a whole, was somewhat less effective in Italy, and was more effective in Japan in late summer 2022, later than in the other G7 countries.

The remainder of the paper proceeds as follows. In Section \ref{section2} we introduce our main theoretical results: we discuss non-asymptotic bounds for a (potentially heterogeneous) neural network panel estimator based on a quadratic loss function. In Section \ref{section3}, we discuss both methodological and implementation aspects of the proposed methodology. We undertake the modeling and forecasting of new COVID-19 cases, and the assessment of the effectiveness of containment policies, in Section \ref{section4}. Section \ref{conclusion} concludes. We relegate to the online appendix additional forecasting results, data summaries, and further discussion of the prediction evaluation tests used.

\section{Theoretical considerations: the deep neural panel data model}\label{section2}

Let $y_{it}$ be the observation for the $i^{th}$ cross-sectional unit at time
$t$ generated by the following panel data model:
\begin{equation}
E\left(y_{it}|\boldsymbol{x}_{it}\right)=\widetilde{h}_{i}\left(  \boldsymbol{x}_{it}\right)
, \quad i=1,\ldots,N,\;t=1,\ldots,T,\label{mod}%
\end{equation}
where $\{\boldsymbol{x}_{it}\}=\{(x_{t,1},\ldots,x_{t,p})^{\prime}\}$ is a
$p$-dimensional vector of regressors, belonging to unit $i$, and $\widetilde
{h}_{i}(\cdot)$ are unknown functions that will be approximated with neural networks.
Throughout, we abstract from unconditional mean considerations, for
simplicity, by assuming $E(y_{it})=0$. This can be achieved by simple unit-by-unit demeaning of the dependent variable. Therefore, the model we entertain is given by: 
\begin{equation}
y_{it}=\widetilde{h}_{i}\left(  \boldsymbol{x}_{it}\right)+\varepsilon_{it}\label{mod20}, 
\end{equation}
where $\varepsilon_{it}$ is an error term. 
Next, we provide a crucial
decomposition to justify the use of a panel structure. We assume that
$\widetilde{h}_{i}\left(  \boldsymbol{x}_{it}\right)$ can be
decomposed as follows:
\begin{equation}
\widetilde{h}_{i}\left(  \boldsymbol{x}_{it}\right)  =h\left(  \boldsymbol{x}%
_{it}\right)  +h_{i}\left(  \boldsymbol{x}_{it}\right), \label{hh}%
\end{equation}
where the function $h(\cdot)$ is the common component of the model, and is our
main focus of interest, and $h_{i}\left(  \boldsymbol{x}_{it}\right)  $ are
idiosyncratic components that will also be approximated with neural networks.
Assumptions needed for the identification of $h(\cdot)$ will be given below. The main
motivation for this decomposition is the familiar linear heterogeneous panel data model, that takes the form:
\begin{equation}
y_{it}=\boldsymbol{x}_{it}^{\prime}\boldsymbol{\beta}_{i}+\varepsilon
_{it}=\boldsymbol{x}_{it}^{\prime}\boldsymbol{\beta}+\boldsymbol{x}%
_{it}^{\prime}\boldsymbol{\eta}_{i}+\varepsilon_{it},\label{mod2}%
\end{equation}
where $E(\boldsymbol{\eta}_{i}|\boldsymbol{x}_{it},\varepsilon_{it})=0$.
Equation \eqref{mod2} allows coefficients to vary across individual units. We
wish to consider and analyze a nonlinear extension of this heterogeneous panel data
model. 

The next step of our proposal involves approximating $h(\cdot)$ and
$h_{i}(\cdot)$ with neural network functional parameterizations, given by $g\left(
\boldsymbol{\cdot};\boldsymbol{\theta}\right)  $. Here, the functional form is
known up to the parameter vector $\boldsymbol{\theta}$, which
is a vector of ancillary parameters, such as network weights and biases. More
details on the choice of $g\left(\boldsymbol{\cdot};\boldsymbol{\theta}\right)$ and the role of various neural network parameters
will be provided in Section \ref{section3} below. Therefore, we parameterize \eqref{hh} by
proposing the following panel model:
\begin{equation}
y_{it}=g\left(  \boldsymbol{x}_{it};\boldsymbol{\theta}^{0}\right)  +g\left(
\boldsymbol{x}_{it};\boldsymbol{\theta}_{i}^{0}\right)  +\varepsilon
_{it},\label{gg}%
\end{equation}
where $\boldsymbol{\theta}^{0}$and $\boldsymbol{\theta}_{i}^{0}$ denote the values of
the parameters that best approximate $h$ and $h_{i}$, respectively (see \eqref{hh}), in a sense to be defined below. 

It is useful to draw some parallels between
\eqref{mod2} and \eqref{gg}. We note that most multi-layer neural network
architectures have a final linear layer given by:
\begin{equation*}
g\left(  \boldsymbol{x}_{it};\boldsymbol{\theta}^{0}\right)
=\boldsymbol{\theta}_{L}^{0\prime}\boldsymbol{f}\left(  \boldsymbol{x}%
_{it}\right),
\end{equation*}
where $\boldsymbol{f}$ is a vector of known functions that form part of the
neural network architecture and $L$ denotes the number of network layers.
Then, it follows that we have a linear representation, in $\boldsymbol{f}$ and
$\boldsymbol{f}_{i}$, of the form:%
\begin{equation}
y_{it}=\boldsymbol{\theta}_{L}^{0\prime}\boldsymbol{f}\left(  \boldsymbol{x}%
_{it}\right)  +\boldsymbol{\theta}_{i,L}^{0\prime}\boldsymbol{f}_{i}\left(
\boldsymbol{x}_{it}\right)  +\varepsilon_{it},\label{nnlin}%
\end{equation}
which is reminiscent of \eqref{mod2} and thus provides a clear rationale for our
nonlinear extension of it. 

Furthermore, it  provides a rationale for thinking that
$\boldsymbol{\theta}_{i}^{0}$ plays a similar role to the idiosyncratic
coefficients, $\boldsymbol{\eta}_{i}$, of the linear model. Of course, one can
use a different network architecture for the panel and idiosyncratic
components, but for simplicity we keep the same structure. The model above encompasses a
variety of nonlinear specifications. It is also worth emphasizing that the
dimension of the regressor vector could be very large. So it is conceivable that each
$\boldsymbol{x}_{it}$ contains regressors from other cross-sectional units,
allowing for complex nonlinear interactions across units. In the limit,
each unit could have $(\boldsymbol{\boldsymbol{x}}_{1t}%
,\ldots,\boldsymbol{\boldsymbol{x}}_{Nt})$ as the regressor vector.

Next, we consider the conditions needed to identify $h(\cdot)$. We require
certain definitions. First, we define $\boldsymbol{\varepsilon}_{it}\equiv
y_{it}-h\left(  \boldsymbol{x}_{it}\right)  -h_{i}\left(  \boldsymbol{x}%
_{it}\right)  $ and $u_{it}=h_{i}\left(  \boldsymbol{x}_{it}\right)
+\boldsymbol{\varepsilon}_{it}$, where the latter is in analogy to the usual
composite error term for the linear heterogeneous panel model, given as
$\boldsymbol{x}_{it}^{\prime}\boldsymbol{\eta}_{i}+\varepsilon_{it}$. The
assumption below generalizes the usual identification assumption on
$\boldsymbol{\eta}_{i}$ made in linear heterogeneous panel models.

\begin{assumption}
\label{Panel_Assumption} For all $i=1,\ldots,N,\;t=1,\ldots,T$

\begin{enumerate}
\item For some positive constant $C$, we assume that $h(\cdot)$ in \eqref{min} below is bounded, such that $\Vert   h\Vert_{\infty}\leq C.$ $h_i$ is bounded similarly to $h$. 
\item $\{u_{it}\}$ is independent and  bounded across $i$.  \label{ui}

\item $E[u_{it}|h\left(  \boldsymbol{x}_{it}\right)  ]=0.$ \label{mds}
\end{enumerate}
\end{assumption}

This assumption enables separation of $h(\cdot)$ and $h_{i}(\cdot)$ when panel
pooled estimation is carried out. For neural network estimation, much stricter
assumptions will be needed. In particular, the second part of the assumption
is justified in view of our later assumption that $h(\cdot)$ and $h_{i}(\cdot)$ can be well
approximated by neural network architectures and the linear aspect of neural
networks discussed in \eqref{nnlin} as it is similar, in functionality, to
assuming that $E(\boldsymbol{\eta}_{i}|\boldsymbol{x}_{it})=0$.

Next we align our discussion with \cite{farrell2021deep}. The
overall goal of neural network estimation in \cite{farrell2021deep} is
to estimate an unknown smooth function $h(\cdot)$ that maps covariates,
$\boldsymbol{X}$, to an outcome ($T\times N$) matrix $\boldsymbol{Y}$, by minimizing a loss
function $g_{\ast}\left(  \boldsymbol{Y},\boldsymbol{X};\boldsymbol{\theta
}\right)  $ with respect to the parameterization $\boldsymbol{\theta}$ of a
neural network function $g\left( \boldsymbol{\cdot}; \boldsymbol{\theta}^{0}\right)  $.
Formally,
\begin{equation}
h=\underset{\boldsymbol{\theta}}{\arg\min}\,E\left[  g_{\ast}\left(
\boldsymbol{Y},\boldsymbol{X};\boldsymbol{\theta}\right)  \right]
.\label{min}%
\end{equation}
This is a minimization of a population quantity and assumes that the true
function $h(\cdot)$ is the unique solution of \eqref{min}. Note that while
\cite{farrell2021deep} do not specify a true function $h(\cdot)$, we take a further step and assume that the true functions in \eqref{hh} coincide with the
unique solutions of \eqref{min}. We do not specify \ $\boldsymbol{Y}$
and $\boldsymbol{X}$ further, since we will apply this general estimation strategy both
to get a panel-based estimate of $h(\cdot)$ and estimates of $h_{i}(\cdot)$ via unit-specific estimation.

For now, we present further sufficient general conditions on $h(\cdot)$ and $g_{\ast}\left(  \boldsymbol{Y},\boldsymbol{X};\boldsymbol{\theta}\right)$ in order for our results to hold. We require the following assumptions:
\begin{assumption}
\label{smoothness} For some constant $C_{g_{\ast}}>0$, we assume that
$g_{\ast}\left(  \boldsymbol{Y},\boldsymbol{X};\boldsymbol{\theta}\right)$
satisfies:%
\begin{equation}
\sup_{\boldsymbol{Y},\boldsymbol{X}}\left\vert g_{\ast}\left(  \boldsymbol{Y}%
,\boldsymbol{X};\boldsymbol{\theta}_{1}\right)  -g_{\ast}\left(
\boldsymbol{Y},\boldsymbol{X};\boldsymbol{\theta}_{2}\right)  \right\vert \leq
C_{g_{\ast}}\left\Vert \boldsymbol{\theta}_{1}-\boldsymbol{\theta}%
_{2}\right\Vert ,\text{ for Frobenius norm }\left\Vert \boldsymbol{\cdot}%
\right\Vert .\nonumber
\end{equation}

\end{assumption}

\begin{assumption}
\label{smoothnes} Consider a H\"{o}lder space $\,\mathcal{W}^{b, \infty
}([-1,1]^{d})$, with $b=1,2,\ldots$, where $\mathcal{W}^{b, \infty}%
([-1,1]^{d})$ is the space of functions on $[-1, 1]^{d}$ in $L^{\infty} $,
along with their weak derivatives. Recall $h$ in \eqref{min}, we assume that
$h$ lies within $\mathcal{W}^{b, \infty}([-1,1]^{d})$, with a norm in
$\mathcal{W}^{b, \infty}([-1,1]^{d})$:
\begin{equation}
\left\Vert h \right\Vert _{\mathcal{W}^{b, \infty}([-1,1]^{d})} =
\max_{\boldsymbol{a}:| \boldsymbol{a} | \leq b} \mathrm{ess} \sup_{x\in[-1,
1]^{d}} \left|  D^{a} h(x) \right| ,
\end{equation}
where $\boldsymbol{a} = (a_{1}, a_{2}, \ldots, a_{d}) \in[-1, 1]^{d}$, $|
\boldsymbol{a}| = a_{1} + a_{2}+\cdots+a_{d}$ and $D^{a} h$ is the
corresponding weak derivative.
\end{assumption}

\begin{remark}
In Assumption \ref{smoothnes} we state a smoothness assumption following
existing theoretical results; see, for example, \cite{farrell2021deep} and
\cite{YAROTSKY2017103, yarotsky2018optimal}. A more detailed discussion of
H\"{o}lder-Sobolev and Besov spaces is available in
\cite{gine2021mathematical}. Assumption \ref{smoothnes} holds for both $h(\cdot)$ and
$h_{i}(\cdot)$, $i=1, \ldots,N.$
\end{remark}

In our case, and in what follows, we specialize the general framework above by
using a squared error loss function that, for the panel setting, becomes:
\[
g_{\ast}\left(\boldsymbol{Y},\boldsymbol{X};\boldsymbol{\theta}\right)=\frac{1}{NT}\sum
_{i=1}^{N}\sum_{t=1}^{T}\left(  {y}_{it}-g(\boldsymbol{x}_{it}%
;\boldsymbol{\theta})\right)  ^{2}.
\]
In our analysis we use feed-forward neural networks architectures with rectified linear unit (ReLU)
activation functions and weights that are unbounded following
\cite{farrell2021deep} and the discussion below. Such networks approximate
smooth functions well, as shown in \cite{YAROTSKY2017103,
yarotsky2018optimal}.

A further assumption is required on the processes $\{
\boldsymbol{x}_{t, k} \}$ and $\{ {\varepsilon}_{it} \}$, for some $k=1,\ldots,
p$, where $p$ is the number of covariates.

\begin{assumption}
\label{Er_cov}
We assume the following:
\begin{enumerate}
\item The rows of $\boldsymbol{X}_{t}$ are \emph{i.i.d.} realizations from a
Gaussian distribution whose $p$-dimensional inner product matrix
$\boldsymbol{\Sigma}$  has a strictly positive minimum eigenvalue, such that
$\Lambda_{\min}^{2} > 0$ and $\Lambda_{\min}^{-2} = O(1). $

\item The rows of the error term $\varepsilon_{it}$ are are \emph{i.i.d.}
realizations from a Gaussian distribution, such that $\varepsilon_{it}\sim
N(0,\sigma_{\varepsilon}I_{N})$.

\item $\varepsilon_{it}$ and $\boldsymbol{x}_{it}$ are mutually independent.
\end{enumerate}
\end{assumption}

\begin{remark}
Assumption \ref{Er_cov} states that the covariate process and the errors are
continuous and have all their existing moments, while being mutually
independent. These strict assumptions are standard in the neural network
literature and useful in order to continue with the analysis on a more
simplified basis$.$ These assumptions are strict. But it is
reasonable to conjecture that similar results to those given below would hold
under weaker conditions.
\end{remark}

Having rewritten the loss function, as squared error loss, with respect to
the re-parameterized panel model in \eqref{gg}, we construct a pooled-type
nonlinear estimator, ${\boldsymbol{\widehat{\theta}}}$, such that:
\begin{equation}
\boldsymbol{\widehat{\theta}}=\underset{\boldsymbol{\theta}}{\arg\min} \, g_{\ast
}\left(\boldsymbol{Y},\boldsymbol{X};\boldsymbol{\theta}%
\right)=\underset{_{\boldsymbol{\theta}\in\mathbb{R}^{d}}}{\arg\min} \, \frac{1}{NT}%
\sum_{i=1}^{N}\sum_{t=1}^{T}\left[y_{it}-g(\boldsymbol{x}_{it}%
;\boldsymbol{\theta})\right]  ^{2},\label{argmin1}
\end{equation}
which obeys Assumption \ref{smoothness}. Therefore, our estimator of $h\left(
\boldsymbol{x}_{it}\right)  $ is given by $g(\boldsymbol{x}_{it}%
;\boldsymbol{\widehat{\theta}})$. Then, we proceed to estimate $h_{i}\left(
\boldsymbol{x}_{it}\right)  $ by $g(\boldsymbol{x}_{it}%
;\boldsymbol{\widehat{\theta}_{i}})$, where $\boldsymbol{\widehat{\theta}_{i}}$
is given by:%
\begin{equation}
\boldsymbol{\widehat{\theta}}_{i}=
\underset{\boldsymbol{\theta}_{i} \in \mathbb{R}^{d}}{\arg\min} \,
\frac{1}{T}\sum_{t=1}^{T}\left[  y_{it}-g(\boldsymbol{x}%
_{it};\boldsymbol{\widehat{\theta}})-g(\boldsymbol{x}_{it};\boldsymbol{\theta
}_{i})\right]  ^{2},\label{argmin2}%
\end{equation}
for each $i$, given $\boldsymbol{\widehat{\theta}}$ from \eqref{argmin1}. \

Next, we argue that the estimation in \eqref{argmin1} can effectively separate
$h\left(  \boldsymbol{x}_{it}\right)  $ from $h_{i}\left(  \boldsymbol{x}%
_{it}\right)  $ and that the unit-wise second step estimation in
\eqref{argmin2} can retrieve $h_{i}\left(  \boldsymbol{x}_{it}\right)  $. We
do this by noting the following. Consider the loss function for
$\boldsymbol{\widehat{\theta}}$ in \eqref{argmin1}. We have:%
\begin{align}
\frac{1}{NT}\sum_{i=1}^{N}\sum_{t=1}^{T}\left[  y_{it}-g(\boldsymbol{x}%
_{it};\boldsymbol{\theta})\right]  ^{2}  & =\frac{1}{NT}\sum_{i=1}^{N}%
\sum_{t=1}^{T}\left[  \left(  h\left(  \boldsymbol{x}_{it}\right)
-g(\boldsymbol{x}_{it};\boldsymbol{ \theta})\right)  +h_{i}\left(
\boldsymbol{x}_{it}\right)  +\varepsilon_{it}\right]  ^{2}\nonumber\\
& =\frac{1}{NT}\sum_{i=1}^{N}\sum_{t=1}^{T}\left[  \left(  h\left(
\boldsymbol{x}_{it}\right)  -g(\boldsymbol{x}_{it};\boldsymbol{ \theta
})\right)  \right]  ^{2}+\frac{1}{NT}\sum_{i=1}^{N}\sum_{t=1}^{T}%
\varepsilon_{it}^{2}\nonumber\\
& +\frac{1}{NT}\sum_{i=1}^{N}\sum_{t=1}^{T}h_{i}\left(  \boldsymbol{x}%
_{it}\right)  ^{2}+\frac{1}{NT}\sum_{i=1}^{N}\sum_{t=1}^{T}\left(  h\left(
\boldsymbol{x}_{it}\right)  -g(\boldsymbol{x}_{it};\boldsymbol{\theta
})\right)  h_{i}\left(  \boldsymbol{x}_{it}\right)  \nonumber\\
& +\frac{1}{NT}\sum_{i=1}^{N}\sum_{t=1}^{T}\left(  h\left(  \boldsymbol{x}%
_{it}\right)  -g(\boldsymbol{x}_{it};\boldsymbol{\theta})\right)
\varepsilon_{it}+\frac{1}{NT}\sum_{i=1}^{N}\sum_{t=1}^{T}h_{i}\left(
\boldsymbol{x}_{it}\right)  \varepsilon_{it}=\sum_{j=1}^{6}A_{j}%
.\label{decomp}%
\end{align}
Under Assumptions \ref{Panel_Assumption}--\ref{Er_cov}, terms $A_{2}$ and
$A_{3}$ converge in probability to positive limits, while $A_{5}$ and $A_{6}$
converge, in probability to zero, and in fact, furthermore, both are
$O_{p}((NT)^{-1/2})$. Additionally, under \eqref{mds}, $A_{4}$ is $O_{p}(N^{-1/2}%
)$. Then it immediately follows that \ the loss function is minimized when
$\boldsymbol{\theta=\theta}^{0}$, in view of our identification assumption in
\eqref{min}. It therefore follows that $\boldsymbol{\widehat{\theta}%
}\rightarrow^{p}\boldsymbol{\theta}^{0}$ and $g(\boldsymbol{x}_{it}%
;\boldsymbol{\widehat{\theta}})\rightarrow^{p}g(\boldsymbol{x}_{it}%
;\boldsymbol{\theta}^{0})=h\left(  \boldsymbol{x}_{it}\right)  $. This proves
that the best pooled panel neural network approximation coincides with the
true panel function.

Next, we  can consider a closely related and, in fact, asymptotically equivalent minimization problem given by:
\begin{equation}\label{ghat}
{g}(\boldsymbol{x}_{it};\boldsymbol{\widehat{\theta}})=
\underset{_{\boldsymbol{\theta}\in\mathbb{R}^{d}}}{\arg\min} \,
\frac{1}{N}\sum_{i=1}^{N}\left[
\frac{1}{T}\sum_{t=1}^{T}y_{it}-\frac{1}{T}\sum_{t=1}^{T}g(\boldsymbol{x}%
_{it};\boldsymbol{\theta})\right]  ^{2}=
\underset{_{\boldsymbol{\theta}\in\mathbb{R}^{d}}}{\arg\min} \,
\frac{1}{N}\sum_{i=1}^{N}\left[  \bar{y}_{i}-\bar{g}%
_{i}(\boldsymbol{\theta})\right]  ^{2}%
\end{equation}
and the associated model with a composite error is:%
\begin{equation}
\bar{y}_{i}=\bar{g}_{i}(\boldsymbol{\theta})+u_{i}, \quad i=1, \ldots, N, 
\end{equation}
where:
\begin{equation}
u_{i}=\frac{1}{T}\sum_{t=1}^{T}g(\boldsymbol{x}_{it};\boldsymbol{\theta}%
_{i})+\frac{1}{T}\sum_{t=1}^{T}\varepsilon_{it}=\bar{g}_{i}%
(\boldsymbol{\theta}_{i})+\bar{\varepsilon}_{i}.
\end{equation}
Note that $u_{i}$ obeys Assumption \ref{Panel_Assumption}.\ref{ui}. Moreover,
this setting corresponds to that of Theorem 1 in \cite{farrell2021deep}
enabling the use of the rates derived in this theorem. This analysis is summarized and extended in the following proposition:
\begin{proposition} 
\label{proposition} 
Suppose Assumptions \ref{Panel_Assumption}--\ref{Er_cov} hold.
Let $g(\boldsymbol{x}_{it}%
;\boldsymbol{\widehat{\theta}})$ be the deep network estimator defined in
\eqref{ghat}. Then, for some $\psi<1/2$, the following holds:
\begin{equation}
\sup_{i,t}\left\Vert g(\boldsymbol{x}_{it};\boldsymbol{\widehat{\theta}%
})-h\left(  \boldsymbol{x}_{it}\right)  \right\Vert _{2}^{2}=O_{P}(N^{-\psi
}).\label{prediction}%
\end{equation}
\end{proposition}

The proof of Proposition \ref{proposition} follows from the proof of Theorem
1 in \cite{farrell2021deep}, using the arguments made above the Proposition to recast our panel framework into the one of \cite{farrell2021deep}, by  separately identifying  $h$ and $h_i$.
In Proposition \ref{proposition}, we use the results from Theorem 1 of
\cite{farrell2021deep} to obtain an asymptotic rate of convergence for the
error in \eqref{prediction}. It is clear that this rate of convergence is not
optimal, since $\psi<1/2$. We have provided a simplified result compared to
Theorem 1 of \cite{farrell2021deep}. Refinements related to factors, such as
the depth and width of the neural network used, can be obtained. These are also
discussed  in Theorem 6 of \cite{bartlett2019nearly} and in Lemma 6 of
\cite{farrell2021deep}. Fast convergence of \eqref{prediction} depends on the
trade-off between the number of neurons and layers, and more specifically on
the parameterization of their relationship,  that controls the approximating
power of the network. 

We note that, in addition, one can obtain consistency for $\boldsymbol{\widehat{\theta}}%
_{i}$ by minimizing the loss over $\boldsymbol{\theta}_{i}$: $L_{i}=\frac{1}%
{T}\sum_{t=1}^{T}[y_{it}-g(\boldsymbol{x}_{it};\boldsymbol{\widehat{\theta}%
})-g(\boldsymbol{x}_{it};\boldsymbol{\theta}_{i})]^{2}.$ Given the rate in
Proposition \ref{proposition}, it immediately follows that $g\left(
\boldsymbol{x}_{it};\boldsymbol{\theta}_{i}^{0}\right)  $ can be consistently
estimated at rate $T^{-\psi}$, as long as $T=o(N^{\xi})$ for some $\xi<1$, given that then the uniform rate in Proposition 1 is faster than $T^{-\psi}$.
\begin{remark}
Before concluding, it is of interest to  consider whether an idiosyncratic component, $g(  \boldsymbol{x}%
_{it};\boldsymbol{\theta}_{i}^{0})  $, is needed, in addition to the
common component, $g(  \boldsymbol{x}_{it};\boldsymbol{\theta}%
^{0})  $. This could be tested by a nonlinear version of a
poolability test. One way to proceed is by fitting only the common component
and then determining whether the residuals, $\widehat{u}_{it}=y_{it}-g(
\boldsymbol{x}_{it};\boldsymbol{\widehat{\theta}})  $, can be further
explained by unit-wise neural network regressions. Again, one way to do this is
by constructing unit-wise $R^{2}$ statistics. This is intuitive, if we recall
the quasi linear representation given by \eqref{nnlin}. One can regress
$\widehat{u}_{it}$ on 
$\boldsymbol{f}_{i}(  \boldsymbol{x}_{it})  $ to obtain such
$R^{2}$ statistics. Then the null hypothesis that $\widetilde{h}%
_{i}\left(  \boldsymbol{x}_{it}\right)  =h(  \boldsymbol{x}_{it})
$, can be tested using the test statistic:
\[
P=\frac{1}{\widehat{\sigma}\sqrt{N}}\sum_{i=1}^{N}\left(  TR_{i}^{2}-m\right),
\]
where $\widehat{\sigma}^{2}=\frac{1}{N}\sum_{i=1}^{N}\left(  TR_{i}^{2}-m\right)
^{2}$ and an appropriate centering factor, $m$, needs to be chosen. This could
be the dimension of $\boldsymbol{f}_{i}\left(  \boldsymbol{x}_{it}\right)  $, although care
needs to be taken, given that  $\boldsymbol{f}_{i}\left(  \boldsymbol{x}_{it}\right)  $
will contain estimated parameters. One way to resolve this issue may be to
estimate the neural networks over a different time period to that
used to run the unit-wise regressions of $\widehat{u}_{it}$ on 
$\boldsymbol{f}_{i}(  \boldsymbol{x}_{it})  $. Then, under our assumptions,
including that of cross sectional independence, and for an appropriate choice
of $m$, $P$ is asymptotically standard normal under the null hypothesis. 
Further exploration of this test is of interest. However, a full and rigorous analysis is beyond the scope of the current paper. 
\end{remark}
\section{Implementation considerations} \label{section3}
In this section we provide details on implementation of the proposed nonlinear estimators. 
First, we summarize the overall neural network construction which relates to the choice of the neural network's architecture and can be summarized by the functional parameterization $g\left(\boldsymbol{\cdot};\boldsymbol{\theta}\right)$,  used in the approximation of $h(\cdot)$. We limit our attention to the construction of   $g\left(\boldsymbol{\cdot};\boldsymbol{\theta}\right)$, since it directly applies to $g\left(\boldsymbol{\cdot};\boldsymbol{\theta_i}\right)$. Then we illustrate how regularization can be applied in the context of the proposed estimators. Finally, we discuss both the cross-validation exercise used to select the different parameters and hyperparameters of the corresponding network and optimization algorithm.

\subsection{Neural network construction}\label{NNconstruction}
We focus on the construction of the \emph{feed-forward neural network}  functional parameterization,  $g\left(\boldsymbol{\cdot};\boldsymbol{\theta}\right)$, used to approximate $h(\cdot)$ in Section \ref{section2}. The feed-forward architecture consists of: an input layer, where the covariates are introduced given an initial set of weights to the inner (hidden) part of the network; the hidden layers, where a number of computational nodes are collected in each hidden layer and nonlinear transformations on the (weighted) covariates occur; and the output layer that gives the final predictions and a choice for the activation function $\sigma(x):\mathbb{R}\rightarrow\mathbb{R}$ that is
applied element-wise. The architecture is feed-forward, since in each of the hidden layers there exist several interconnected neurons that allow information to flow from one layer to the other, but only in one direction. The connections between layers correspond to weights.

We use $L$ to define the total number of hidden layers and $M^{(l)}$, $l=1, \ldots, L$ to define the total number of neurons at the $l^{th}$ layer. $L$ and $M^{(l)}$ are measures for the depth and width of the neural network, respectively.
 We use the ReLU activation function, $\sigma_l(\boldsymbol{X}_{t}):=\max(\boldsymbol{X}_{t}, 0)$, where $\boldsymbol{X}_{t}$ is a $N\times p $ matrix of characteristics for $ t=1,\ldots, T$; $l=1, \ldots,L-1$  and a linear activation function for $l=L$. The activation functions are applied elementwise.  To explain the exact computation of the outcome of the \emph{feed-forward neural network},  we focus on the pooled-type estimator in  \eqref{argmin1}.  We assume that the widths (the number of neurons), $M^{(l)}$, and depth (the number of hidden layers), $L$, of the network are  constant positive numbers.
 
 Each of the neurons undergoes a computation similar to the linear combination received  in each hidden layer $l$: $\boldsymbol{g}^{(l)} = \sigma_{l} (\boldsymbol{g}^{(l-1)}\boldsymbol{W}^{(l)'} + \boldsymbol{b}^{(l)'}) $, while the final output of the network is $\boldsymbol{g}^{(L)} =\boldsymbol{g}^{(L-1)}\boldsymbol{W}^{(L)'} + \boldsymbol{b}^{(L)'}$ and  $\boldsymbol{g}^{(0)} = \boldsymbol{X}_{t}$.  We can then define  for some $t=1,\ldots, T,$ $g\left(\boldsymbol{\cdot};\boldsymbol{\theta}\right)$ as: 
 \begin{equation}\label{convolution}
 g\left(\boldsymbol{X}_{t}; \boldsymbol{\theta} \right) = \left(\sigma_{L}%\left( \sigma_{L-1}\left(
 \cdots \sigma_{2}\left(  \sigma_{1}\left(\boldsymbol{X}_t\boldsymbol{W}^{(1)'} + \boldsymbol{b}^{(1)'}\right)\boldsymbol{W}^{(2)'} +\boldsymbol{b}^{(2)'}\right)  \cdots\right) %\right)\boldsymbol{W}^{(L-1)'}+\boldsymbol{b}^{(L-1)'}\right)
 \boldsymbol{W}^{(L)'} + \boldsymbol{b}^{(L)'},
 \end{equation} 
where $\boldsymbol{W}^{(l)} $ is a $M^{(l)} \times M^{(l-1)}$ matrix of weights, $\boldsymbol{b}^{(l)}$ is a $M^{(l)} \times N$ matrix of biases at layer $l$, with $\boldsymbol{b}^{(1)} = \boldsymbol{0}$. Notice that  at $l=1$, the dimensions of  $\boldsymbol{W}^{(1)} $ are $M^{(1)}\times p$ and of  $\boldsymbol{b}^{(1)}$  are $M^{(1)} \times N$. At the final layer, that is, at $l=L$, the dimensions of $\boldsymbol{W}^{(L)} $ are $1 \times M^{(L-1)}$ and of $\boldsymbol{b}^{(L)}$  are $1 \times N$.

 Note that throughout the paper we use $\boldsymbol{\theta}$ to denote a stacked vector containing all ancillary trainable parameters affiliated with the network estimation, as defined below:
 \begin{equation}\label{parameters}
 \boldsymbol{\theta} = \left( \mathrm{vec}\left(\boldsymbol{W}^{(1)'}\right), \ldots, \mathrm{vec}\left(\boldsymbol{W}^{(L)'}\right), \mathrm{vec}\left(\boldsymbol{b}^{(1)'}\right), \mathrm{vec}\left(\boldsymbol{b}^{(2)'}\right), \ldots, \boldsymbol{b}^{(L)'}  \right)'.
 \end{equation}
  We define the overall number of parameters as $d=|\boldsymbol{\theta}|$.   The optimization of the neural network proceeds in a forward fashion (from the input layer, that is, $l=1$, to the output $l=L$) and layer-by-layer through an optimizer, for example, a version of stochastic gradient descent (SGD), where the gradients of the parameters $(\boldsymbol{W}^{(l)}, \boldsymbol{b}^{(l)})$ are calculated through back-propagation (using the chain-rule) to train the network. 
 \begin{remark}
The exact (composition) structure described in \eqref{convolution} holds for a
subclass of \emph{feed-forward neural networks}, specifically that one that
refers to fully connected layers (the one being consecutive to the other) but
has no other connections. Each layer has a number of hidden units that are of
the same order of magnitude. This architecture is the most commonly used in
empirical research, and is often referred to as a
\emph{Multi-layer Perceptron} (MLP). Furthermore, the exact structure in
\eqref{convolution} does not hold generally for any \emph{feed-forward neural
network}.
\end{remark}

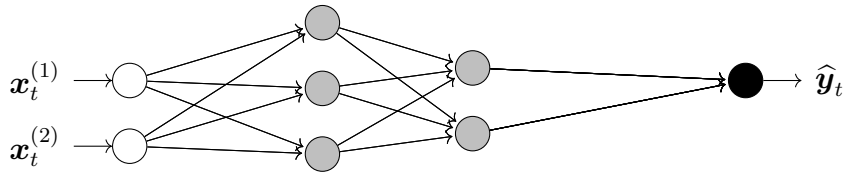
\begin{figure}[!t]
\centering
\begin{tikzpicture}[shorten >=1pt,->, draw=black,
		node distance = 4mm and 34mm,
		start chain = going below,
		every pin edge/.style = {<-,shorten <=0pt},
		neuron/.style = {circle, fill=#1,
			minimum size=13pt, inner sep=0pt,
			on chain},
		annot/.style = {text width=4em, align=center}
		]
		
		% Draw the input layer nodes
		\foreach \i in {1, 2}
		\node[draw=black, neuron=white!50,
		pin=180: $\boldsymbol{x}_{t}^{(\i)}$] (I-\i)  {}; %  {$x_{\i}$};
		
		% Draw the hidden layer nodes
		\node[draw=black, neuron=gray!50,
		above right=6mm and 24mm of I-1.center] (H-1)   {}; %  {$x_{1}$};
		\foreach \i [count=\j from 1] in {2,3}
		\node[draw=black,neuron=gray!50,
		below=of H-\j]      (H-\i)   {}; % {$x_{\i}$};
		
		% Draw the hidden layer nodes
		\node[draw=black,neuron=gray!50,
		above right=0mm and 44mm of I-1.center] (G-1)  {}; %   {$x_{1}$};
		\foreach \i [count=\j from 1] in {2}
		\node[draw=black,neuron=gray!50,
		below=of G-\j]      (G-\i)  {}; %  {$x_{\i}$};
		
		%
		% Draw the output layer node
		\node[draw=black,neuron=black,
		pin= {[pin edge=->]0:$\boldsymbol{\widehat{y}}_{t}$ },
		right=of I-1 -| H-2 -| G-2]  (O-1) {}; %  {$x_{1}$};
		\foreach \i [count=\j from 1] in {1}
		%             \node[neuron=red!50,
		%             pin= {[pin edge=->]0:Output \#\j},
		%             below=of O-\j]        (O-\i)  {$x_{\i}$};
		% Connect input nodes with hidden nodes and
		%  hiden nodes with output nodes with the output layer
		\foreach \i in {1,...,2}
		\foreach \j in {1,...,3}
		\foreach \k in {1,...,2}
		{
			\path (I-\i) edge (H-\j)
			(H-\j)  edge (G-\k)
			(G-\k) edge (O-1);
		}
	\end{tikzpicture}
\caption{Illustration of a \emph{feed-forward neural network} with two input matrices
$(\boldsymbol{x}_{t}^{(1)}, \boldsymbol{x}_{t}^{(2)})^{\prime}$, two layers, $L=2$, 5
nodes, $M=5$, eighteen connections, $W=14$, and one (fitted) output
${\protect\boldsymbol{\widehat{y}}}_{t}.$ The inputs are illustrated with a white
circle, the neurons with grey circles, the output with a black circle. }%
\label{nn_fig}%
\end{figure}

The specific choice of the network architecture is crucial and affects the
complexity and the approximating power of $g\left(\boldsymbol{\cdot};\boldsymbol{\theta}\right)$ in \eqref{convolution}.
Our analysis involves primarily theoretical arguments that are widely
applicable in \emph{feed-forward neural networks} when we deal with panel data. We present an example of a
\emph{feed-forward neural network}, based on \eqref{convolution}, in Figure \ref{nn_fig}.

The neural network in Figure \ref{nn_fig} consists of two inputs
$\boldsymbol{X}_{t}\in\mathbb{R}^{N\times p}$, $\boldsymbol{X}_{t} = ( \boldsymbol{x}_{t}^{(1)}%
, \boldsymbol{x}_{t}^{(2)} )$, in particular, $p=2$,
 where $\boldsymbol{x}_{t}^{(j)}$ is a $N\times 1$ vector of one characteristic at  $t=1,\ldots, T$ for some $j=1,2$,  and one fitted output $\boldsymbol{\widehat{y}}_{t}$. Between the inputs and output
$(\boldsymbol{X}_{t}, \boldsymbol{\widehat{y}}_{t})^{\prime}$, are $M$ hidden
computational nodes/neurons, in particular $M=5$. The neurons are connected
directly forming an acyclic graph which specifies a fixed architecture.\footnote{ The network in Figure \ref{nn_fig} can be used to optimize \eqref{argmin1} and also \eqref{argmin2}, but now used for each $i= 1,\ldots,N$. }

Notice that the illustration in Figure \ref{nn_fig} can
correspond to a nonlinear pooled-type estimation of $g( \boldsymbol{X}%
_{t};\boldsymbol{\theta}^{0}) $ in \eqref{gg}, where we use \eqref{argmin1}
to obtain $\boldsymbol{\widehat{\theta}}$,   defined in \eqref{parameters}, with input  $\boldsymbol{X}_{t} =
(\boldsymbol{x}^{(1)}%
_{t}, \ldots, \boldsymbol{x}^{(p)}_{t} )$,
% where
% $\boldsymbol{x}^{(j)}_{t}\in\mathbb{R}^{N\times p}$, $j=1,\ldots,p$ 
and output
$\boldsymbol{\widehat{y}}_{t}\in\mathbb{R}^{N\times 1}. $ The remainder of
\eqref{gg}, $g( \boldsymbol{X}_{t};\boldsymbol{\theta}_{i}^{0}), \;
i=1,\ldots,N, \; t=1,\ldots, T $ can be described conceptually as the heterogeneous component,
that differs cross-sectionally. One can obtain the estimate of this
heterogeneous component following the same steps as those used to obtain $g(
\boldsymbol{x}_{it};\boldsymbol{\widehat{\theta}}) $, with the only difference
that now the \emph{feed-forward neural network} is estimated
unit-wise, similar to the logic of a fixed effects estimator for
linear panel data models.
\subsection{Implementation and regularization} \label{section3.2}
In this section we discuss some operational implementation aspects required for the estimation of the panel neural network estimators proposed in Section \ref{section2}. We focus discussion on the following panel estimator,  ${g}(\boldsymbol{x}_{it}; \boldsymbol{\widehat{		\theta}})$, obtained from  the optimization of \eqref{ghat}:
\[
	g(\boldsymbol{x}_{it};\boldsymbol{\widehat{\theta}})=
 \underset{_{\boldsymbol{\theta}\in\mathbb{R}^{d}}}{\arg\min} \,
\frac{1}{N}\sum_{i=1}^{N}\left[ \frac{1}{T}\sum_{t=1}^{T}y_{it}-\frac{1%
	}{T}\sum_{t=1}^{T}g(\boldsymbol{x}_{it};\boldsymbol{\theta })\right]
	^{2}.
\]

This nonlinear panel estimator, and generally neural network estimators,   have many significant advantages over traditional panel models, mainly summarized in their great capacity at approximating highly nonlinear and complicated associations between variables and outstanding forecasting performance; see, for example, the discussion in \cite{goodfellow2016deep} and \cite{gu2020empirical, gu2020autoencoder}.  In order to be able to minimize \eqref{ghat} and obtain a feasible solution for the panel estimator ${g}(\boldsymbol{x}_{it}; \boldsymbol{\widehat{		\theta}})$, we need to choose the overall architecture of the neural network. Following the discussion above, this reduces to choices for the total number of layers $L$, total number of neurons $M^{(l)}$, at each $l=1,\ldots, L$ layers, a loss function $g_{\ast}(\boldsymbol{y},\boldsymbol{X;\theta})$, which in this paper is taken to be the MSE loss, an updating rule for the weights (learning rate, $\gamma$) during optimization, and the optimization algorithm itself, typically taken to be some variant of SGD.

However, neural networks tend to overfit,  which can lead to a severe deterioration in their (forecasting) performance. A common empirical solution to this is to impose a penalty on the trainable parameters of the neural network, $\boldsymbol{\theta}$. The penalized estimator based on the \textsc{LASSO} is obtained as the solution to the following minimization problem:
\[
g(\boldsymbol{x}_{it};\boldsymbol{\widehat{\theta}})^{\textsc{LASSO}}=
 \underset{_{\boldsymbol{\theta}\in\mathbb{R}^{d}}}{\arg\min} \,
\frac{1}{N}\sum_{i=1}^{N}\left[ \frac{1}{T}\sum_{t=1}^{T}y_{it}-\frac{1%
	}{T}\sum_{t=1}^{T}g(\boldsymbol{x}_{it};\boldsymbol{\theta })\right]
	^{2} + \lambda \left \Vert \boldsymbol{\theta} \right \Vert_{1},
\]
where $\lambda$ is the regularization parameter. Note that while explicit regularization improves empirical solutions of neural networks estimators
under low signal-to-noise ratios, its role is not clear theoretically, since there are cases where simpler SGD solutions present similar solutions; see, for example,  \cite{zhang2021understanding}.   
Other  commonly used regularization techniques frequently employed empirically to assist in the estimation of neural networks, relate to batch normalization,  early stopping, and dropout. We succinctly discuss batch normalization below, given its importance because of the cross-sectional aspect of our estimator. We refer the reader to \cite{gu2020empirical} for a detailed discussion of early stopping and dropout.

Batch normalization, proposed by \cite{ioffe2015batch}, is a technique used to control the variability of the covariates across different regions of the network and datasets. It is used to address the issue of internal covariate shift, where inputs of hidden layers may follow different distributions than their counterparts in the validation sample. This is a prevalent issue when fitting, in particular, deep neural networks. Effectively, batch normalization cross-sectionally demeans and standardizes the variance of the batch inputs.

\begin{remark}\label{remark_double_descen}
In this paper, we consider both penalized and non-penalized estimation. While using the latter might seem problematic due to the large number of parameters that needs to be estimated, we find, in our empirical work, that this is not necessarily the case. This is not surprising. Recent work in the statistical and machine learning literature highlights what is known as the double descent effect. For linear regressions, this relates to the use of generalized inverses to construct least squares estimators, when the number of variables, $p$, exceeds the number of observations, $T.$ Such estimators work better either when $p$ is small (and standard matrix inversion can be used) or when $p$ is much larger than $T.$  Then the quality of the performance of an estimator is implicitly measured in terms of the ``bias-variance trade-off,'' where an optimal performance resides at the lowest reported bias and variance of the corresponding model (either linear or nonlinear).  While it is widely accepted that the ``bias-variance trade-off'' function resembles a U-shaped curve, it has been observed, for example see  \cite{belkin2019reconciling} and  \cite{hastie2022surprises},  that beyond the interpolation limit the test loss descends again, hence the name ``double-descent.''  To understand why, note that such estimators implicitly impose penalization by using generalized inverses and so choose the parameter vector with the smallest norm, among all admissible vectors; see \cite{hastie2022surprises}.  So once $p$ is much larger than $T$ such a selection becomes more consequential, as many more candidate vectors are admissible. This linear effect is also present for neural network estimation given the connection to linear models highlighted in Section \ref{section2}, as discussed in detail in \cite{hastie2022surprises} and \cite{kelly2022virtue}.
\end{remark}

\subsection{Cross-validation}\label{section:cv}

The cross-validation (CV) scheme consists of choices on the overall architecture of the neural network:  the total number of layers ($L$), neurons ($M$), the learning rate ($\gamma$) of SGD,
the batch size, dropout rate, level of regularization $(\lambda)$, and a choice on the activation functions. 

Regarding the choice on the activation functions, we use ReLU for the hidden layers and a linear function for the output layer.  We tune the learning rate of the optimizer, $\gamma$, from
five discrete values in the interval $\left[0.01, 0.001\right]$. We tune the depth and width of the neural networks using the following grids, $[1,3,5,10,15]$ and [5, 10, 15, 20, 30], respectively. Hence the choice between deep or shallow learning is completely data-driven, as it is selected from the CV scheme. We set the batch size to 14. For the tuning of the regularization parameter, $\lambda$, used for \textsc{LASSO} penalisation, we use the following grid $c\sqrt{\log{p}/NT}$, where $c=[0.001, 0.01, 0.1, 0.5, 1, 5, 10]$. We also use  dropout regularization, where the dropout probability is up to $10$ percent; see, for example, \cite{gu2020empirical}.

To select the trainable parameters, $\boldsymbol{\theta}$, and the hyperparameters discussed above, we follow \cite{gu2020empirical, gu2020autoencoder} and divide our data into three disjoint time periods that maintain the temporal ordering of the data: the \textit{training} sub-sample, which is used to estimate the parameters of the model, $\boldsymbol{\theta}$, given a specific set of hyperparameters; the \textit{validation} sub-sample, which is used to tune the different hyperparameters given $\boldsymbol{\widehat{\theta}}$ from the \textit{training} sub-sample\footnote{Note that while $\boldsymbol{\widehat{\theta}}$ is used in the tuning of the hyperparameters, it is only estimated at the \textit{training} sub-sample.}; and, finally, the \textit{testing} sub-sample which is truly out-of-sample and is used to evaluate our nonlinear models' forecasting performance. As discussed in detail below, our forecasting exercise is recursive, based on an expanding window size. Hence, at each expanding window, we need to use the train-validation-split of the sample and estimate the relevant parameters and tune the hyperparameters. At each expanding window, let $T^\ast$ denote the total sample size for the specific window, then the \textit{training} sub-sample consists of $\lfloor 0.8T^\ast \rfloor$, the validation sub-sample consists of $\lfloor 0.2 T^\ast \rfloor - c$, and finally, the testing sub-sample consists of 7, 14, or 21 observations depending on the forecast horizon, $h$, respectively. $c$ is chosen so that the testing sub-sample always has $h$ observations and $\lfloor \cdot \rfloor$ stands for the floor function. 

\subsection{Optimization}
The estimation of neural networks is generally a computational cumbersome optimization problem due to nonlinearities and non-convexities. The most commonly used solution utilizes SGD to train a neural network. SGD uses a batch of a specific size, that is, a small subset of the data at each \textcolor{black}{epoch (iteration)} of the optimization to evaluate the gradient, to alleviate the computation hurdle. The step of the derivative at each epoch is controlled by the
learning rate, $\gamma$. We use the adaptive moment estimation algorithm (ADAM) proposed by \cite{kingma2014adam}\footnote{ADAM is using estimates for the first and second moments of the gradient to calculate the learning rate.}, which is a more efficient version of SGD.  
Finally, we set the number of epochs to $5,000$ and use  early stopping following \cite{gu2020empirical}  to mitigate potential overfitting.

\section{Empirical analysis: forecasting new COVID-19 cases} \label{section4}
In this section, after introducing the data, we examine the predictive ability of the proposed model(s) for forecasting the daily path of new COVID-19 cases across the G7 countries. We compare the forecasting results from our new models against two restricted alternatives: a neural network without a cross-sectional dimension and a linear panel data VAR (PVAR). Comparison against these alternatives lets us examine the importance of firstly modeling the panel dimension and secondly of allowing for nonlinearities. To assess the out-of-sample Granger causality of pandemic-induced lockdown policies on the spread of COVID-19, we compare the forecasting performance of our models with and without measures of the stringency of government-imposed containment and lockdown policies. Such (non-pharmaceutical) policies were differentially adopted by many countries from March 2020, including the G7, to reduce the spread of COVID-19. Then, we discuss how partial derivatives can be used to help interpret the output of the deep panel models. They can be used to help assess the efficacy of the different containment policy measures taken by individual countries to contain the spread of COVID-19. 
		
\subsection{The COVID-19 data and the Oxford stringency index}

Our interest is modeling and forecasting, at a daily frequency, reports of new COVID-19 cases per 100K of the population over the sample April 2020 through December 2022 for the G7 countries. We source these data from the World Health Organization coronavirus dashboard. 

As $\boldsymbol{x}_{it}$ variables, for each country, $i$, at day, $t$, we consider a set of 7 lagged COVID-19 related indicators, as well as lags of new cases per-100K (our $y_{it}$ variable). For parsimony, we confine attention to lags at 7, 14, 21, and 28 days. These 7 variables, plus lags of the dependent variable, may all have explanatory power for $y_{it}$. The 7 variables (all reported per 100K of the population) comprise: new
deaths, the reproduction rate, new tests, the share of COVID-19 tests that are positive measured as a rolling 7-day average (this is the inverse of tests per case), the number of people vaccinated, the number of people fully vaccinated, and the number of total boosters. \cite{knutson2022estimating}, \cite{mathieu2021global}, and \cite{CAPORALE202277} also consider such COVID-19 related variables, given that they all likely relate (contemporaneously or at a lag) to the number of new COVID-19 cases.

To assess the role of containment policies in explaining and forecasting the spread of new COVID-19 cases, we then consider specifications that augment the aforementioned set of $\boldsymbol{x}_{it}$ variables by adding in a measure or measures of the stringency of the government response to COVID-19. Specifically, we use the government response stringency index, as compiled by the Oxford Coronavirus Government Response Tracker (OxCGRT). This index is a composite measure based on 9 response indicators, namely: school closures, workplace closures, the cancellation of public events, restriction on gatherings, public transport closing, requirements to stay at home, movement restriction, restrictions on international travel, and public information campaigns. Throughout the pandemic the Oxford stringency index was a widely consulted measure of policy. Since the Oxford index is an aggregation of 9 indicators, with the weights subjectively chosen by Oxford, we also experiment with forecasting when the underlying 9 disaggregates enter individually into our models, so that, in effect, we objectively use the data to weight the disaggregates. Note that we always consider the lagged effects of policy changes on new COVID-19 cases, mitigating endogeneity concerns that, for example, stricter lockdown policies follow increases in new COVID-19 cases. 

Throughout, $t$ corresponds to a day and we use  a trailing seven-day rolling average to smooth the data. The cross-sectional dimension of our panel is $p=36$ when we consider the aggregate stringency index (as published by Oxford) and $p=68$ when we consider the disaggregated stringency index. We further
follow the literature (see, for example, \cite{gu2020autoencoder}) and rank-normalize all of our variables into the 
$[0, 1]$ interval as follows:
\[
\widetilde{\boldsymbol{x}}_i = \frac{\boldsymbol{x}_i - \min(\boldsymbol{x}_i)}{\max(\boldsymbol{x}_i)-\min(\boldsymbol{x}_i)}, \quad i=1,\ldots, N.
\]
This normalization minimizes the influence of severely outlying observations stemming from covariate distributions that may have significant departures from normality, a common feature of COVID-19 data, especially at the beginning of the pandemic. 

\begin{figure}[!ht]		
\centering
	\includegraphics[width=\linewidth]{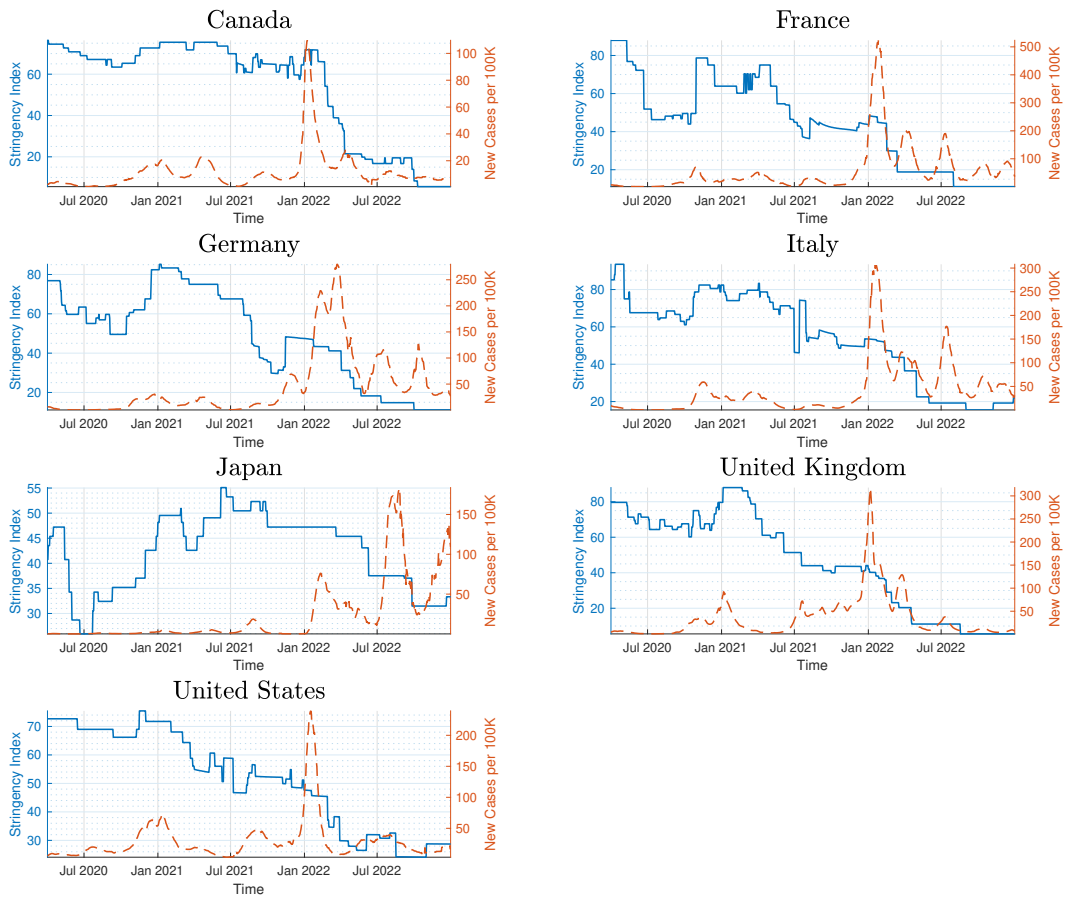}
	\caption{The Oxford stringency index and new COVID-19 cases per 100K of the  population}	\label{strinfig}
\end{figure}

The online Data Appendix provides additional data details. Figure \ref{strinfig} presents the aggregate stringency index and plots new COVID-19 cases per 100K of the population through our sample period. This figure shows that there are apparent commonalities across countries, both in the stringency of policy and the evolution of new COVID-19 cases. But there are differences too, with Japan standing out as having looser containment policies than the other countries during mid-2020  and then experiencing a later spike in new COVID-19 cases in summer 2022. Thus, it remains an empirical question whether forecasting new COVID-19 cases is improved by pooling information across countries.

\subsubsection{Out-of-sample forecasting design}\label{fdesign}

We recursively produce forecasts of $y_{it}$ -- new COVID-19 cases -- by estimating our set of models using expanding estimation windows and evaluate these forecasts over the out-of-sample period February 6, 2021 through December 24, 2022. Given (\ref{gg}), the $h-$day-ahead forecast of new COVID-19 cases per 100K is:
\begin{equation}\label{fore}
	\widehat{y}_{i,t+h} |\mathcal{F}_t  =  	\widehat{g}\left( \boldsymbol{x}_{it};\boldsymbol{\theta }^{\star}\right) +	\widehat{g}\left(\boldsymbol{x}_{it};\boldsymbol{\theta }_{i}^{\star}\right),
\end{equation}
where $	\widehat{g}\left( \boldsymbol{x}_{it+h};\boldsymbol{\theta }^{\star}\right)$ denotes the corresponding fit of the pooled network, $\widehat{g}\left( 
\boldsymbol{x}_{it+h};\boldsymbol{\theta }_{i}^{\star}\right) $ denotes the unit-by-unit fit of the  network, and 	$\widehat{y}_{i,t+h} |\mathcal{F}_t$ denotes the deep idiosyncratic forecast. $ \mathcal{F}_t$ denotes the information set up to time $t$, for some $t=1,\ldots, T$,   $ \boldsymbol{\theta }^{\star}$ denotes the optimal weights obtained from the CV for the deep pooled model, and $\boldsymbol{\theta }_{i}^{\star}$  denotes the optimal weights obtained from the CV for the deep idiosyncratic model.  We compare this forecast against, what we call, the ``deep pooled'' forecast that sets $\widehat{g}\left( 
	\boldsymbol{x}_{it};\boldsymbol{\theta }_{i}^{\star}\right) =0$. 

We recursively compute $h=7$, $h=14$, and $h=21$ day-ahead forecasts using an expanding estimation window (relating $y_{i,t+h}$ to $\boldsymbol{x_{it}}$, as per (\ref{fore})). To ease the computational burden, given that we re-estimate the model and use CV (as discussed in Section \ref{section3}) at each window, we increase the size of the estimation windows in increments of 7 days. We now summarize how estimation and forecasting works for $h=7$ (forecasting at the longer horizons proceeds analogously): We first estimate our models using daily data from April 1, 2020 through January 30, 2021 ($T^{0}=305$) and produce forecasts 7 days-ahead. Then we estimate from April 1, 2020 through February 6, 2021 ($T^{1}=312$)  and again produce forecasts 7 days-ahead. We carry on this process until we finally estimate our models over the sample April 1, 2020 through December 17, 2021 ($T^{700}=991$) producing forecasts 7 days-ahead. This results in an out-of-sample sample size of $700$ days. We do not consider forecasting earlier than 7 days-ahead, given that the incubation period of COVID-19 is typically around one week, so that we should not expect policy changes to have effects within one week. During the first wave of the pandemic, many governments revised their policy measures to restrict the virus once a week, which also helps rationalize our choice of forecast horizons. Forecasts for longer horizons, $h$, are obtained similarly.

To test if and how our proposed deep neural network panel data models confer forecasting gains, we compare them against two benchmarks that switch off firstly panel (cross-country) interactions and secondly nonlinear effects. We do so by estimating: (i) a ``deep time-series'' model that is identical to our deep neural network panel data model but is estimated separately for each country; and (ii) a panel VAR (PVAR) model that does allow for cross-country interactions, but assumes linearity in terms of how $x_{it}$ affects $y_{it+h}$. Testing our model against these two special cases isolates whether it is allowing for cross-country interaction and/or for nonlinearity that is advantageous. 

We follow \cite{canova2009panel} and specify the $i^{th}$ equation of the PVAR with $q$ lags as:
\begin{equation}
	y_{it}=  A_{1i}\boldsymbol{Y}_{t-1}+\cdots+ A_{qi}\boldsymbol{Y}_{t-q}+  \epsilon_{it},   \epsilon_{it} \sim \text{i.i.d.}N\left(0, \sigma_{i}\right), \label{pvar}
\end{equation}
where $A_{ji}$ for $j=1,\ldots, q$ are coefficient matrices, we have dropped the intercept for notational simplicity, $\boldsymbol{Y}_t= ({z}_{1t}', \ldots, {z}_{Nt}')'$, and $z_{it}= ({y}_{it}, \boldsymbol{x}_{it})'$.   We set $q= 28$. We estimate the PVAR by OLS and compute $h$-day-ahead forecasts of $y_{it+h}$ from (\ref{pvar}) via iteration.

\subsubsection{Forecast evaluation}
In this section we evaluate the forecasting performance of the proposed nonlinear panel estimator(s) relative to the two benchmark models, namely, the linear PVAR(28), and the deep time-series neural network. We then  examine whether the inclusion of policy related variables affects forecast accuracy. Specifically, to test for out-of-sample Granger causality of the policy measures adopted by governments to contain the spread of COVID-19, we compare the forecast accuracy of all of our models with and without the aggregate and disaggregate Oxford stringency indexes.

%We use two different tests to examine, first, the statistical significance of the forecasts and second any potential forecast instability, that we will discuss below. 
We evaluate the accuracy of the forecasts of new COVID-19 cases using the \text{root mean squared forecast
	error}  (RMSE):  
\begin{equation*}
	\textrm{RMSE}_{i} = \sqrt{\frac{1}{T}{\sum_{t=1}^{T}\left ({y}_{i, t+h}- \widehat{y}_{i,t+h} |\mathcal{F}_t  \right)^2}}, \quad i=1,\ldots, N.
\end{equation*}

We use the \cite{Dieb95} {(DM)} test to test whether differences in forecast accuracy across models are statistically significant. We follow \cite{harvey1997testing} and use their small-sample adjustment.

Table \ref{tbl7app} compares the accuracy of our two deep nonlinear models against the two benchmarks when we do not include the stringency-based measures of policy and instead focus on predicting new COVID cases using lags of new COVID cases and the other 7 COVID-related measures. The results are striking. Both deep nonlinear panel models provide significant forecasting gains over both the linear PVAR(28) model and the deep time-series neural network at all three forecast horizons. This shows the importance of both the panel dimension and nonlinearites in forecasting the daily path of new COVID-19 cases across the G7 countries. Of the two deep models, the deep pooled estimator delivers, for all 7 countries, more accurate forecasts than the deep idiosyncratic model. Simpler models often work better when forecasting and this appears to be the case here too: allowing for additional country-specific effects in our deep pooled model hinders out-of-sample forecasting performance. Tables \ref{dm1}-\ref{dm3} in the online appendix show that the forecasting gains, of the deep models against the time series model, are statistically significant. This evidences that the gains from modeling and forecasting new COVID-19 cases come from pooling data (in a nonlinear manner) across the G7 countries.

\begin{table}[!ht]
\centering
\resizebox{!}{!}{%
\begin{tabular}{@{}lccccccc@{}}
\toprule
                   & Canada & France & Germany & Italy & Japan & UK    & US    \\ \midrule 
                   & \multicolumn{5}{c}{$h=7$} \\ \midrule 
Deep pooled        & 0.073  & 0.095  & 0.112   & 0.074 & 0.135 & 0.089 & 0.065 \\
Deep idiosyncratic & 0.125  & 0.136  & 0.141   & 0.111 & 0.163 & 0.118 & 0.108 \\
Deep time-series   & 0.321  & 0.326  & 0.365   & 0.311 & 0.420 & 0.313 & 0.256 \\
PVAR(28)           & 0.242  & 0.323  & 0.221   & 0.205 & 0.373 & 0.246 & 0.282 \\ \midrule
& \multicolumn{5}{c}{$h=14$} \\ \midrule
Deep pooled        & 0.091  & 0.118  & 0.117   & 0.088 & 0.138 & 0.101          & 0.075         \\
Deep idiosyncratic & 0.138  & 0.153  & 0.146   & 0.122 & 0.167 & 0.130          & 0.115         \\
Deep time-series   & 0.301  & 0.297  & 0.345   & 0.293 & 0.391 & 0.305          & 0.256         \\
PVAR(28)           & 0.233  & 0.308  & 0.204   & 0.194 & 0.343 & 0.246          & 0.278\\ \midrule
& \multicolumn{5}{c}{$h=21$} \\ \midrule
Deep pooled        & 0.117 & 0.131 & 0.131 & 0.106 & 0.157 & 0.117 & 0.097 \\
Deep idiosyncratic & 0.160 & 0.162 & 0.152 & 0.143 & 0.182 & 0.148 & 0.130 \\
Deep time-series   & 0.292 & 0.284 & 0.337 & 0.289 & 0.388 & 0.299 & 0.265 \\
PVAR(28)           & 0.231 & 0.294 & 0.198 & 0.190 & 0.318 & 0.249 & 0.280\\
\bottomrule
\end{tabular}%
}
\caption{RMSE statistics for the 7, 14, and 21 day-ahead forecasts of new COVID-19 cases from the 4 models without policy-related variables over the sample February 6, 2021 through December 24, 2022.  The reported models are: Deep pooled: $ \widehat{g}\left( \boldsymbol{x}_{it+h};\boldsymbol{\theta }^{\star}\right) $; Deep idiosyncratic: $\widehat{g}\left( \boldsymbol{x}_{it+h};\boldsymbol{\theta }^{\star}\right) +\widehat{g}\left( 
	\boldsymbol{x}_{it+h};\boldsymbol{\theta }_{i}^{\star}\right) $; Deep time-series: $ \widehat{g}\left( \boldsymbol{x}_{t+h};\boldsymbol{\theta }_{TS}^{\star}\right) $; and the PVAR(28) model.  $\boldsymbol{\theta}^{\star}$,  $\boldsymbol{\theta}_{i}^{\star}$, and $\boldsymbol{\theta}_{TS}^{\star}$ are obtained via out-of-sample CV. 
		   }\label{tbl7app}
\end{table}

We next test whether the containment or lockdown policies, imposed at the national-level, help forecast new COVID-19 cases. If the policies were effective, conditioning on them should deliver more accurate forecasts. Table \ref{tb4a} presents the relative RMSE ratios for each of the four forecasting models when estimating including and excluding the aggregate stringency index. We see that focusing on the deep models, given their higher accuracy as seen in Table \ref{tbl7app}, policy as measured by the aggregate stringency index was only effective in France and Japan at 7 days. In the other 5 countries, the RMSE ratios are greater than unity, indicating that better forecasts of new COVID-19 cases are made without the stringency index. Interestingly, for the less accurate deep time-series and PVAR models, policy appears to have been more effective. But consistent with it taking time for policy changes to affect the path of the pandemic,  Table \ref{tb4a} shows that after an additional two weeks policy was effective in all G7 countries, except Italy and the US.  

Table \ref{tb5a} then tests whether the Oxford stringency data have more value-added when forecasting if we let the models decide how much weight to attach to each of the 9 components (policy levers) in the aggregate stringency index. The fact that the RMSE ratios, for the preferred deep models, are now less than unity across all 7 countries indicates that policy was effective after all: but it is important to let the data determine what policies matter in which country. Table \ref{tb5a} indicates that at $h=7$ days policy was least effective in Canada and Italy, as while policy interventions still affect new COVID-19 cases, unlike in the other G7 countries, these effects are not statistically significant. However, again demonstrating that policy changes take time to have impact, policy has a larger effect after another week (at $h=14$ days), as in both Canada and Italy the relative RMSE ratios are lower at 14 days than at 7 days.

\begin{table}[!ht]
\centering
\resizebox{!}{!}{%
\begin{tabular}{@{}llllllll@{}}
\toprule
		& Canada & France & Germany & Italy & Japan & UK    & US    \\ \midrule 
                 &  \multicolumn{5}{c}{$h=7$} \\ \midrule 
Deep pooled        & 1.084  & 0.928  & 1.033   & 1.229 & 0.878 & 1.043 & 1.321 \\
Deep idiosyncratic & 1.015  & 0.897  & 1.038   & 1.152 & 0.836 & 1.309 & 1.329 \\
Deep time-series   & 0.847  & 0.946  & 1.076   & 0.881 & 1.044 & 0.949 & 1.040 \\
PVAR(28)           & $0.875^{***}$  & $0.868^{***}$ & 0.933  & $0.850^{*}$ & $0.792^{**}$ & $0.880^{*}$ & $0.596^{**}$\\ 
\midrule
& \multicolumn{5}{c}{$h=14$} \\ \midrule
Deep pooled        & 1.012  & $0.907^{*}$   & 0.896   & 1.154 & 0.861 & 0.952          & 1.225         \\
Deep idiosyncratic & 0.967  & $0.868^{*}$  & 0.913   & 1.087 & 0.841 & 1.210          & 1.191         \\
Deep time-series   & 0.910  & 0.915  & 1.085   & 0.920 & 1.045 & 0.920          & 0.960         \\
PVAR(28)           & $0.880^{***}$  & $0.856^{***}$  & 0.930   & $0.855^{*}$ & $0.773^{**}$ & $0.891^{*}$ & $0.603^{**}$  \\ \midrule
& \multicolumn{5}{c}{$h=21$} \\ \midrule
Deep pooled        & 0.953 & $0.879^{*}$ & $0.864^{*}$ & 1.117 & $0.855^{*}$ & 0.957 & 1.078 \\
Deep idiosyncratic & 0.903 & $0.838^{**}$ & 0.874 & $0.989^{*}$ & 0.850 & 1.124 & 1.059 \\
Deep time-series   & 0.951 & 0.878 & 1.094 & 0.923 & 1.033 & 0.936 & 0.947 \\
PVAR(28)           & $0.887^{***}$ & $0.839^{***}$ & 0.928 & $0.854^{*}$ & $0.794^{**}$ & $0.889^{*}$ & $0.607^{***}$ \\
\bottomrule	
\end{tabular}
}
	\caption{RMSE ratios, comparing the forecast accuracy of each respective model with and without the aggregate Oxford stringency index at 7, 14, and 21 days-ahead.  Ratios $<1$ indicate superior  predictive ability for the model with  the stringency index.  For a description of the 4 forecasting models, see the notes to Table \ref{tbl7app}.  $*$, $**$, and $***$ denote  rejection of the null hypothesis of equality of forecast mean squared errors with and without the aggregate Oxford stringency index at the 10\%, 5\%, and 1\% levels of significance, respectively, using the modified \cite{Dieb95}  test with the \cite{harvey1997testing} adjustment.}
 \label{tb4a}
\end{table}

\begin{table}[!ht]
\centering
\resizebox{!}{!}{%
\begin{tabular}{@{}llllllll@{}}
\toprule
		& Canada & France & Germany & Italy & Japan & UK    & US    \\ \midrule 
             &      \multicolumn{5}{c}{$h=7$} \\ \midrule 
Deep pooled        & 0.941  & $0.834^{**}$  & $0.852^{**}$   & 0.848 & $0.852^{*}$ & $0.842^{**}$ & $0.845^{**}$ \\
Deep idiosyncratic & 0.866  & 0.861  & 0.913   & 0.861 & 0.875 & 1.059 & 0.947 \\
Deep time-series   & 0.906  & 1.063  & 1.071   & 0.909 & 1.067 & 0.901 & 1.075 \\
PVAR(28)           & 0.840  & $0.614^{***}$  & 0.992   & 0.791 & 0.807 & 0.850 & 0.739 \\ 
\midrule
& \multicolumn{5}{c}{$h=14$} \\ \midrule
Deep pooled        & 0.921  & 0.893  & $0.868^{**}$    & $0.843^{**}$  & $0.860^{*}$  & $0.887^{*}$  & $0.908^{*}$          \\
Deep idiosyncratic & $0.796^{*}$   & 0.900  & 0.955   & $0.822^{*}$  & 0.877 & 1.021          & 0.925         \\
Deep time-series   & 0.934  & 1.079  & 1.066   & 0.917 & 1.096 & 0.890          & 1.045         \\
PVAR(28)           & 0.826  & $0.579^{***}$   & 1.043   & 0.783 & 0.802 & 0.840          & $0.732^{*}$     \\ \midrule
& \multicolumn{5}{c}{$h=21$} \\ \midrule
Deep pooled        & 0.900 & $0.884^{**}$ & $0.845^{**}$ & $0.825^{***}$ & $0.843^{**}$ & 0.931 & $0.887^{***}$ \\
Deep idiosyncratic & $0.772^{**}$ & 0.869 & 0.991 & $0.787^{**}$ & 0.891 & 0.953 & 0.864 \\
Deep time-series   & 0.994 & 1.090 & 1.118 & 0.919 & 1.084 & 0.906 & 1.009 \\
PVAR(28)           & 0.830 & $0.557^{***}$ & 1.078 & 0.785 & 0.846 & 0.834 & $0.730^{**}$    \\
\bottomrule
	\end{tabular}
}
	\caption{RMSE ratios, comparing the forecast accuracy of each respective model with and without the disaggregate Oxford stringency index at 7, 14, and 21 days-ahead. Ratios $<1$ indicate superior  predictive ability for the model with  the stringency index.  For a description of the 4 forecasting models, see the notes to Table \ref{tbl7app}.  $*$, $**$, and
			$***$ denote  rejection of the null hypothesis of equality of forecast mean squared errors with and without the disaggregate Oxford stringency index at the 10\%, 5\%, and 1\% levels of significance, respectively, using the modified \cite{Dieb95}  test with the \cite{harvey1997testing} adjustment.}\label{tb5a}
\end{table}

In the online appendix, we provide additional checks on the forecasting performance of our models. We show that the forecasting gains from the models conditioning on the disaggregate stringency index are often stronger in the first half of our out-of-sample window, when in absolute terms the forecasting errors were higher as COVID-19 infection rates were higher and more volatile. Analysis also indicates that the gains of our deep pooled models, over the linear PVAR model, were higher during these earlier waves of COVID-19. This is consistent with the pandemic exhibiting highly nonlinear features in its earlier waves, before vaccinations and other immunities helped restrain the spread of COVID-19. The fluctuation test of \cite{giacomini2010forecast} is used to show that policy in Italy and Japan proved to be effective later than in the other G7 countries: it is only by the fall of 2022 that we see policy having a marked effect on forecast accuracy. We also present results with the lasso penalization and discuss the observed double descent pattern that our deep models forecast better without any penalty.

\subsection{Policy effectiveness}

A common critique of ML algorithms is their putative trade-off between accuracy and interpretability. The output of a highly complicated ML model, such as a deep neural network of the sort we consider, may fit the data well in-sample and even, as we find, out-of-sample. But the model itself is often hard to interpret. In this section, we illustrate how the use of partial derivatives provides one way to assess the impact of covariates. We focus on examination of the effects of changes in policy, as measured by the aggregate and the disaggregate stringency indexes, on the transmission of new COVID-19 cases.

The use of partial derivatives to interpret model output is, of course, common practice in econometrics,  ranging from the simple linear regression model to impulse response analysis. In this section, we show how partial derivatives can be used in deep neural networks to interpret highly nonlinear relationships between covariates and the dependent variable.\footnote{We prefer the use of partial derivatives over Shapley additive explanation values, as proposed by \cite{lundberg2017unified}, since 
derivatives tend to be less noisy (see, for example, \cite{chronopoulos2023forecasting}) and computationally less expensive to compute.  Perhaps, though, the biggest
disadvantage is the set of implicit assumptions, used in the operational construction of Shapley values.
A major one is the assumption that inputs are statistically independent. This is discussed in \cite{AIonshap},
who also discuss solutions. However, these are computationally intensive,
potentially still quite poor approximations, and not appropriate for large
sets of inputs. While partial derivatives (as well as coefficients in linear
models) have similar issues, as discussed in \cite{pes2014a}, these issues are
both more transparent in nature, and, as discussed in \cite{pes2014a}, far
easier to address.}

While our deep neural networks are highly nonlinear, their solution/output via SGD optimization methods, can be treated as differentiable functions, as the majority of activation functions are differentiable. In this paper, we consider the case of  ReLU, which is not differentiable at zero, whereas it is at every other point of $\mathbb{R}$.  From a computational standpoint, the gradient descent, heuristically,  works well enough to treat it as a differentiable function. Furthermore, \cite{goodfellow2016deep} argue that this issue is negligible and machine learning softwares are prone to rounding errors, making them very unlikely to compute the gradient at a singularity point. Note that even in this extreme case, both SGD and ADAM, will  use the right sub-gradient at zero.

Let the matrix of characteristics be denoted $\boldsymbol{X}_t \in \mathbb{R}^{N\times p}$, where $\boldsymbol{X}_t = (\boldsymbol{x}^{(1)}_{t}, \ldots, \boldsymbol{x}^{(p)}_{t})$. Then for some $i=1, \ldots, N$, $j=1, \ldots, p$ and $t=1,
\ldots, T$, the partial derivatives of $g(\boldsymbol{X}_t; \widehat{\theta})$ with respect to the $j^{th}$ characteristic in $\boldsymbol{X}_{t}$ are: 
\begin{equation} %\label{eq22}
	d_{i \, j, \, t} = \frac{\partial{g}\left(\boldsymbol{X}_{t}; \widehat{\boldsymbol{\theta}}\right)}{\partial {x}_{i,\, j, \, t-h}},
 \label{partial}
\end{equation}
where 
 ${g}(\boldsymbol{X}_t; {\boldsymbol{\widehat{\theta}}})$ is the  function (see Section \ref{section2}) that approximates the number of new cases per-$100$K across the $i$ different countries, in our case the G7 countries. We assess the partial derivatives across time since, following \cite{kapetanios2007measuring}, we expect them to vary  due to the inherent nonlinearity of the neural network. 

In our work we present the partial derivatives, defined in \eqref{partial}, without adding confidence bands around them to assess statistical  significance.  The reason for this is that there is currently no rigorous technology in the literature to produce these, especially in the case of  penalized estimation.  However, recent work by \cite{kap2023}  uses a bootstrap approach to  construct confidence bands around partial derivatives. A full modification of this work for use in  panel models is an interesting and promising avenue to proceed, but is left for future research.

\begin{figure}[!t]	
\centering
	\includegraphics[width=\linewidth]{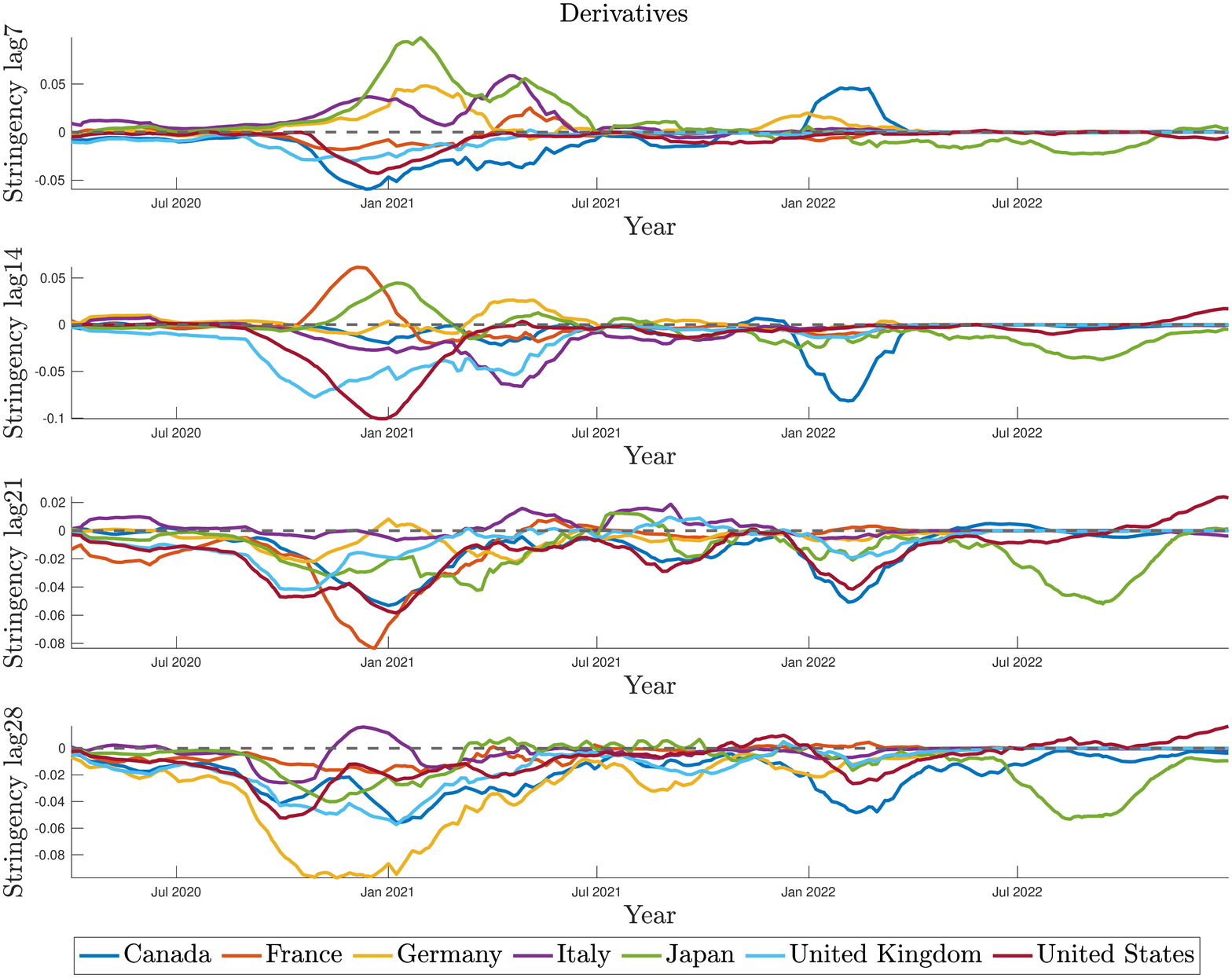}
		\caption{Partial derivatives: The effects of policy (as measured by the Oxford stringency index) on new COVID-19 cases 7, 14, 21, and 28 days after the policy change.}	\label{Derivatives}
\end{figure}	

In Figure \ref{Derivatives}, we present  the partial derivatives with respect to the aggregate stringency index at horizons $h\in \{7, 14, 21, 28 \}.$  Thereby we evaluate the dynamic effectiveness of the stringency policies adopted across the G7 countries.\footnote{
We present the partial derivatives as sixty-day moving averages to smooth out noise.} There are three features that we draw out from Figure \ref{Derivatives}. First, policy is more effective at containing the spread of COVID-19 after 7 days. Stronger and more negative effects of increases in stringency are seen after 7 days. Secondly, with the exception of in Japan, policy was most effective in the late fall of 2021 and in early 2022, at the time of the highly contagious Omicron variant. The dynamic effects of policy are, on average, much weaker in the second half of our sample. This is consistent with higher vaccination rates meaning that from mid-2021 (non-immunization) policies became less effective at restraining the spread of new COVID-19 cases. Thirdly, there is considerable cross-country variation in the effectiveness of policy. As referenced above when summarizing the \cite{giacomini2010forecast} fluctuation tests reported in the online appendix, policy in Japan is again seen in Figure \ref{Derivatives} to have been most effective in late-summer 2022, consistent with COVID-19 cases peaking later in Japan than in the other countries (see Figure \ref{strinfig}). Containment policies in Italy tended, relative to the other countries, to have a more muted effect.

Given the evidence from Table \ref{tb5a} that the disaggregated stringency index confers additional forecasting gains relative to the aggregate index, we next look at the partial derivatives with respect to the 9 components of the Oxford index. This way we aim to shed light on the effectiveness of specific policy measures. We focus on the effects of school and university closings and of workplace closings, since of the 9 components of the Oxford stringency index these tend to be the specific policies associated with the largest marginal effects.  Results for the other policy measures are provided in the online appendix. Given the high degree of correlation between the different policy measures (see online Tables \ref{fig2app}-\ref{fig3app}), we should in any case not over interpret these partial derivatives.

\begin{figure}[!b]		
\centering
	\includegraphics[width=0.75\linewidth]{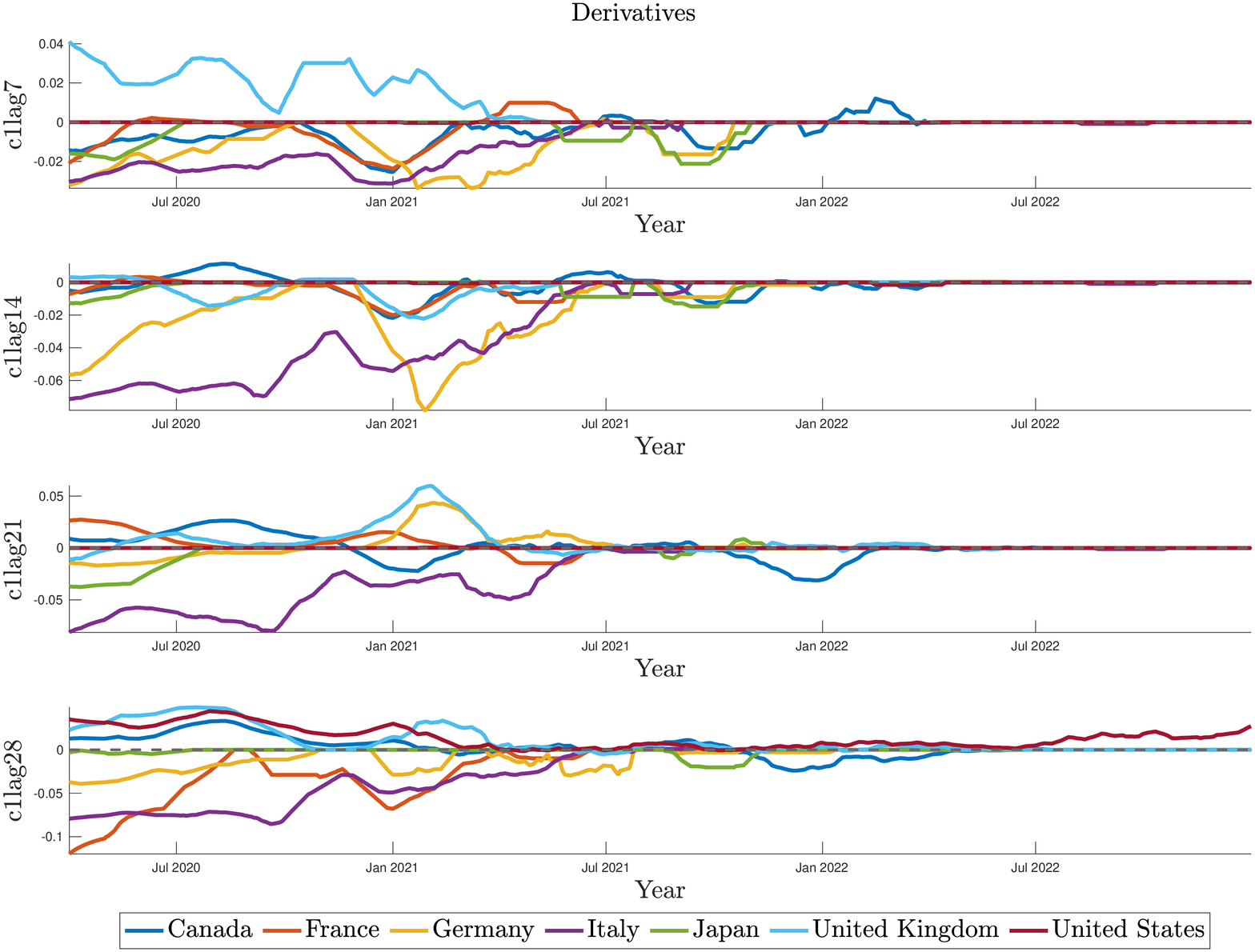}
	\caption{Partial derivatives: The effects of school and university closures on new COVID-19 cases 7, 14, 21, and 28 days after the policy change }	
  \label{d1}
\end{figure}

Figure \ref{d1} shows that over time (as $h$ increases from 7 to 28 days) the effects of school and university closings had an increasingly strong effect. For most countries, as expected, these effects are negative: the closures lead to a fall in new COVID-19 cases. These negative effects are especially strong in Italy. But in the UK, the effects are not so clean cut, with the closings appearing to have a positive effect during the early stages of COVID-19. As in Figure \ref{Derivatives}, we again see evidence across countries that the effects of school and university closures were far more effective prior to January 2022. Thereafter, the effects are much more modest.

\begin{figure}[!b]		
\centering
	\includegraphics[width=0.75\linewidth]{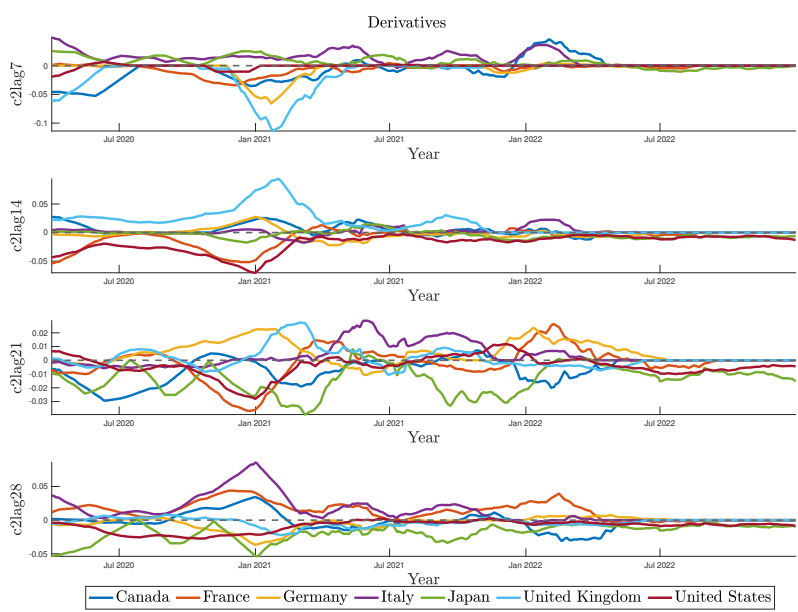}
	\caption{Partial derivatives: The effects of workplace closures on new COVID-19 cases 7, 14, 21, and 28 days after the policy change}	
  \label{d2}
\end{figure}

Turning to Figure \ref{d2}, we see that while workplace closures tended to have a negative effect on COVID-19 soon after the policy change, in particular in Germany and the UK, thereafter the effects are more uncertain and variable across countries. This can be attributed not just to difficulties in isolating the direct effects of one policy change versus another (related) one, but because in the intervening period there were likely additional and perhaps offsetting changes.

\clearpage

		\section{Conclusion} \label{conclusion}
            This paper proposes a nonlinear panel data estimator of the conditional mean based on neural networks. We explore heterogeneity and latent patterns in the cross-section, and derive an estimator to account for these patterns. Furthermore, we provide asymptotic arguments for the proposed methodology building on the work of \cite{farrell2021deep}.

            We use the proposed estimators to forecast, in a simulated out-of-sample experiment, the progression of the COVID-19 pandemic across the G7 countries. We find significant forecasting gains over both linear panel data models and  time series neural networks. Containment or lockdown policies, as instigated at the national-level, are found to have out-of-sample predictive power for the spread of new COVID-19 cases. Using partial derivatives to help interpret the panel neural networks, we find considerable heterogeneity and time-variation in the effectiveness of specific containment policies.

%\newpage
		%\bibliographystyle{econometrica}
		\bibliography{covidkc}

		\clearpage

\appendix
\setcounter{equation}{1}
 \setcounter{page}{1}
\counterwithin{table}{section}
 \counterwithin{figure}{section}
 
\noindent{\LARGE{\textbf{Online Appendix}}}
\section{Data Appendix}

This appendix provides additional details on the dataset used in the empirical analysis in Section \ref{section4} of the main paper.  Specifically, we present the main variables for each country, $i$, that constitute the design matrix $\boldsymbol{X}$; and we provide summary statistics for each variable considered.  We assemble our data set from four publicly available different sources. The stringency index is obtained from the Oxford Coronavirus Government Response Tracker (OxCGRT) (these data can be found at \url{https://www.bsg.ox.ac.uk/research/covid-19-government-response-tracker}).  The daily confirmed COVID-19 cases, which is the ``raw data'' version of our response, as well as the daily confirmed   deaths, were collected from the World Health Organization Coronavirus Dashboard (available at \url{https://covid19.who.int/?mapFilter=cas}). The  official numbers and metrics from governments and health ministries, worldwide, regarding vaccinations were collected from \cite{mathieu2021global}. Lastly, the testing and virus passivity-rates data are from \cite{owidcoronavirus}.

The rapid spread of COVID-19 led countries to take drastic measures to contain the virus and protect their health systems. The OxCGRT data set gathers together a  set of  longitudinal measures of government responses from January 1, 2020.   These measures include school closings, national/international travel restrictions, bans on public gatherings, emergency investments in healthcare facilities, new forms of social welfare provision, contact tracing, among others; see \cite{hale2021global} for more details.    These different measures are then aggregated into one unified measure -- the stringency index -- that records the strictness of policies that primarily restricted people’s behavior, such as via lockdowns.  The index is calculated using all ordinal containment and closure policy indicators, including  an indicator recording public information campaigns.  The higher the value of this index, the stricter the policies adopted.   Table \ref{tbl1app} presents the nine different response indicators underlying the aggregate stringency index.

Figure \ref{fig1app} presents the correlation matrix across new
deaths, the reproduction rate, new tests, the share of COVID-19 tests that are positive measured as a rolling 7-day average (this is the inverse of tests per case), the number of people vaccinated, the number of people fully vaccinated, the number of total boosters, and new cases per-100K.   In Figures \ref{fig2app} -- \ref{fig3app} we present  the correlation matrix of the different variables  for each G7 country.

In Figure \ref{fig1app} a high negative correlation between the new cases per-$100$K and the stringency index and its nine components is observed. This of course makes sense, as it implies that when more cases emerge, the stricter the containment policies adopted. Furthermore, there exist a positive correlation between the stringency index and its nine constituent components.

\begin{table}[!ht]
\centering
\resizebox{\textwidth}{!}{%
\begin{tabular}{@{}lccl@{}}
\toprule
ID & Name                               & Description                                                                                                      & \multicolumn{1}{c}{Coding}                                                                                                                                                                                    \\ \midrule
C1 & c1m\_school\_closing               & Record closings of schools and universities                                                                      & 0 - no measures                                                                                                                                                                                               \\
   &                                    &                                                                                                                  & \begin{tabular}[c]{@{}l@{}}1 - recommend closing or all schools open with alterations resulting in significant differences \\ compared to non-COVID-19 operations\end{tabular}                                \\
   &                                    &                                                                                                                  & 2 - require closing (only some levels or categories, eg just high school, or just public schools)                                                                                                             \\
   &                                    &                                                                                                                  & 3 - require closing all levels                                                                                                                                                                                \\
   &                                    &                                                                                                                  & Blank - no data                                                                                                                                                                                               \\ \midrule
C2 & c2m\_workplace\_closing            & Record closings of workplaces                                                                                    & 0 - no measures                                                                                                                                                                                               \\
   &                                    &                                                                                                                  & \begin{tabular}[c]{@{}l@{}}1 - recommend closing (or recommend work from home) or all businesses open with alterations\\ resulting in significant differences compared to non-Covid-19 operation\end{tabular} \\
   &                                    &                                                                                                                  & 2 - require closing (or work from home) for some sectors or categories of workers                                                                                                                             \\
   &                                    &                                                                                                                  & 3 - require closing (or work from home) for all-but-essential workplaces (eg grocery stores, doctors)                                                                                                         \\
   &                                    &                                                                                                                  & Blank - no data                                                                                                                                                                                               \\ \midrule
C3 & c3m\_cancel\_public\_events        & Record cancelling public events                                                                                  & 0 - no measures                                                                                                                                                                                               \\
   &                                    &                                                                                                                  & 1 - recommend cancelling                                                                                                                                                                                      \\
   &                                    &                                                                                                                  & 2 - require cancelling                                                                                                                                                                                        \\
   &                                    &                                                                                                                  & Blank - no data                                                                                                                                                                                               \\ \midrule
C4 & c4m\_restrictions\_on\_gatherings  & Record limits on gatherings                                                                                      & 0 - no restrictions                                                                                                                                                                                           \\
   &                                    &                                                                                                                  & 1 - restrictions on very large gatherings (the limit is above 1000 people)                                                                                                                                    \\
   &                                    &                                                                                                                  & 2 - restrictions on gatherings between 101-1000 people                                                                                                                                                        \\
   &                                    &                                                                                                                  & 3 - restrictions on gatherings between 11-100 people                                                                                                                                                          \\
   &                                    &                                                                                                                  & 4 - restrictions on gatherings of 10 people or less                                                                                                                                                           \\
   &                                    &                                                                                                                  & Blank - no data                                                                                                                                                                                               \\ \midrule
C5 & c5m\_close\_public\_transport      & Record closing of public transport                                                                               & 0 - no measures                                                                                                                                                                                               \\
   &                                    &                                                                                                                  & 1 - recommend closing (or significantly reduce volume/route/means of transport available)                                                                                                                     \\
   &                                    &                                                                                                                  & 2 - require closing (or prohibit most citizens from using it)                                                                                                                                                 \\
   &                                    &                                                                                                                  & Blank - no data                                                                                                                                                                                               \\ \midrule
C6 & c6m\_stay\_at\_home\_requirements  & \begin{tabular}[c]{@{}c@{}}Record orders to "shelter-in-place"\\  and otherwise confine to the home\end{tabular} & 0 - no measures                                                                                                                                                                                               \\
   &                                    &                                                                                                                  & 1 - recommend not leaving house                                                                                                                                                                               \\
   &                                    &                                                                                                                  & 2 - require not leaving house with exceptions for daily exercise, grocery shopping, and 'essential' trips                                                                                                     \\
   &                                    &                                                                                                                  & \begin{tabular}[c]{@{}l@{}}3 - require not leaving house with minimal exceptions (eg allowed to leave once a week, \\ or only one person can leave at a time, etc)\end{tabular}                               \\
   &                                    &                                                                                                                  & Blank - no data                                                                                                                                                                                               \\ \midrule
C7 & c7m\_movementrestrictions          & \begin{tabular}[c]{@{}c@{}}Record restrictions on internal\\ movement between cities/regions\end{tabular}        & 0 - no measures                                                                                                                                                                                               \\
   &                                    &                                                                                                                  & 1 - recommend not to travel between regions/cities                                                                                                                                                            \\
   &                                    &                                                                                                                  & 2 - internal movement restrictions in place                                                                                                                                                                   \\
   &                                    &                                                                                                                  & Blank - no data                                                                                                                                                                                               \\ \midrule
C8 & c8ev\_internationaltravel          & Record restrictions on international travel.                                                                     & 0 - no restrictions                                                                                                                                                                                           \\
   &                                    & \begin{tabular}[c]{@{}c@{}}Note: this records policy for foreign travellers,\\ not citizens.\end{tabular}        & 1 - screening arrivals                                                                                                                                                                                        \\
   &                                    &                                                                                                                  & 2 - quarantine arrivals from some or all regions                                                                                                                                                              \\
   &                                    &                                                                                                                  & 3 - ban arrivals from some regions                                                                                                                                                                            \\
   &                                    &                                                                                                                  & 4 - ban on all regions or total border closure                                                                                                                                                                \\
   &                                    &                                                                                                                  & Blank - no data                                                                                                                                                                                               \\ \midrule
H1 & h1\_public\_information\_campaigns & Record presence of public info campaigns.                                                                        & 0 - no Covid-19 public information campaign                                                                                                                                                                   \\
   &                                    & Note no differentiated policies reported in this indicator.                                                      & 1 - public officials urging caution about Covid-19                                                                                                                                                            \\
   &                                    &                                                                                                                  & 2- coordinated public information campaign (eg across traditional and social media)                                                                                                                           \\
   &                                    &                                                                                                                  & Blank - no data                                                                                                                                                                                               \\ \bottomrule
\end{tabular}%
}
\caption{Mnemonics for the 9 components of the Oxford stringency index} \label{tbl1app}
\end{table}

\begin{figure}[!ht]		
\centering
	\includegraphics[width=0.95\textwidth]{Figs/corr_heat}
	\caption{ Correlation matrix (pooled across the G7 countries) between the COVID-19 variables and the Oxford stringency variables}\label{fig1app}	
\end{figure}	

\begin{figure}[!ht]	
\centering
\includegraphics[width=0.80\textwidth]{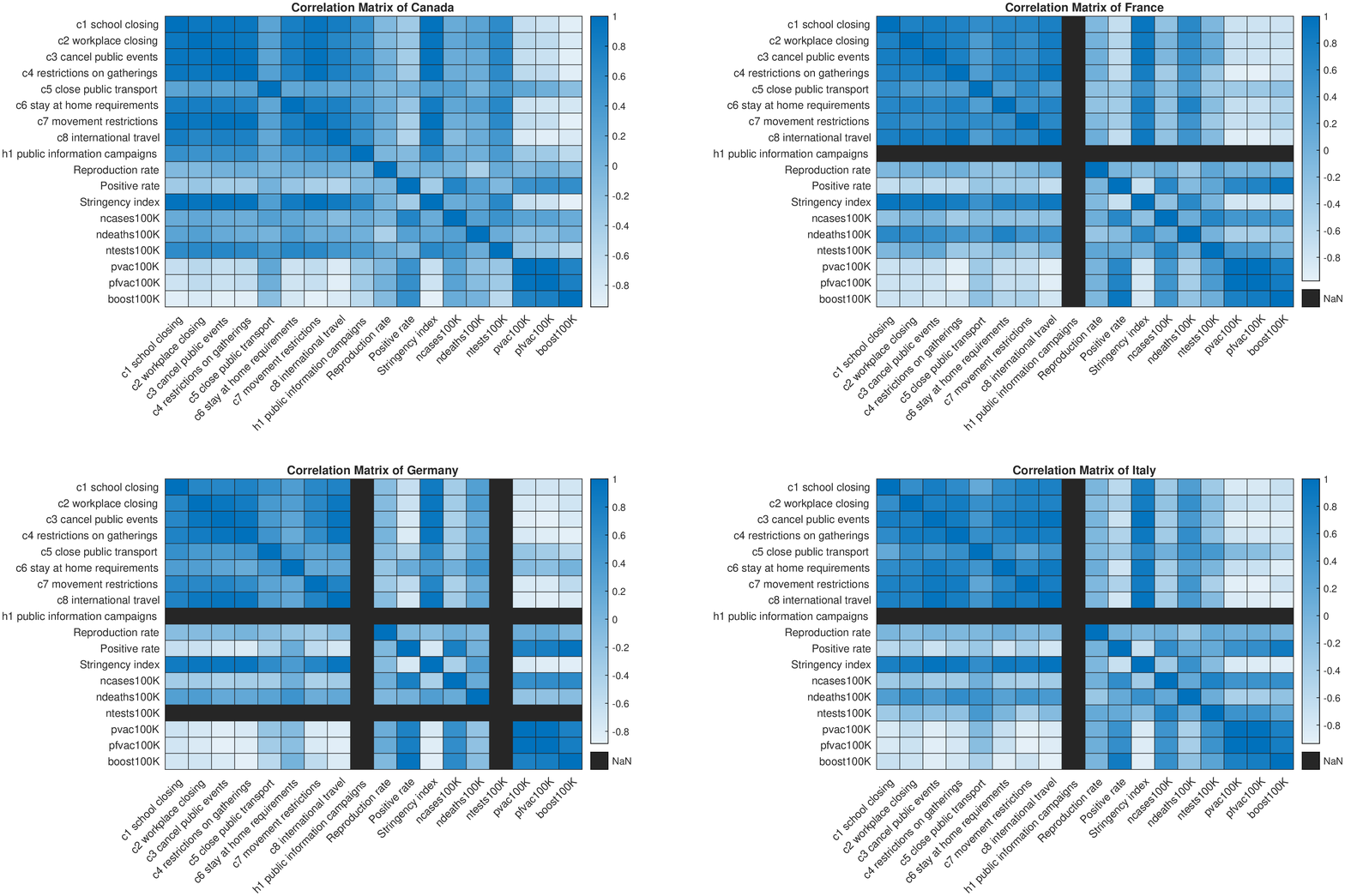}
\caption{ Correlation matrix by country between the COVID-19 variables and the Oxford stringency variables}\label{fig2app}	
\end{figure}	
\clearpage
\begin{figure}[!ht]		
\centering
	\includegraphics[width=0.80\textwidth]{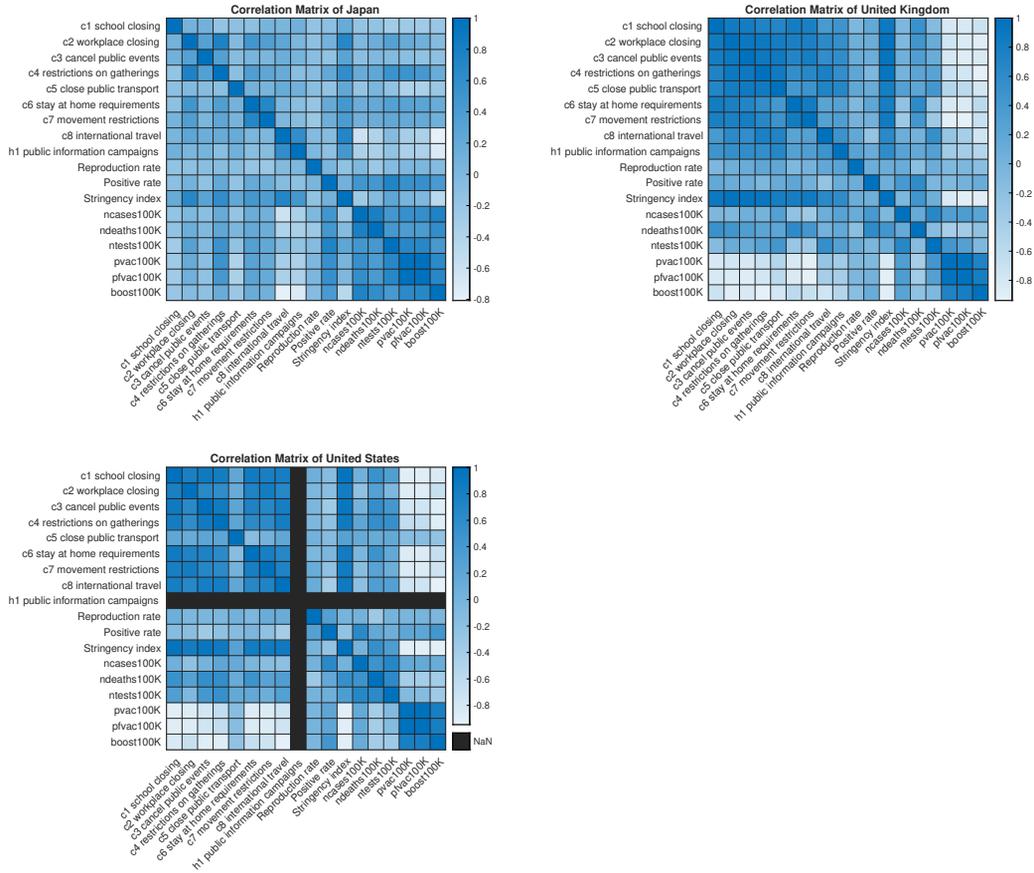}
	\caption{ Correlation matrix by country between the COVID-19 variables and the Oxford stringency variables (cont.)}\label{fig3app}	
\end{figure}	

\section{Additional empirical results}

In this section, we present supplementary results as referenced in the main paper.

\subsection{Diebold-Mariano tests for equal forecast performance against the deep time-series model} 
In this section we present the relative RMSE ratios, in order to compare the forecasting accuracy of each model against the deep time series model, described in more detail in Section \ref{section4}.   Similarly to the main paper, we examine the forecasting ability of models without policy variables specifically the stringency index (see Table \ref{dm1}), including the aggregated stringency index (see Table \ref{dm2}), and including the dissagregated stringency index (see Table \ref{dm3}).
\bigskip
\bigskip
\bigskip
\begin{table}[h]
\centering
\resizebox{!}{!}{%
\begin{tabular}{@{}llllllll@{}}
\toprule
                      & Canada & France & Germany & Italy & Japan & UK    & US    \\ \midrule 
                   & \multicolumn{5}{c}{$h=7$} \\ \midrule 
Deep pooled        & $0.227^{***}$    & $0.293^{***}$    & $0.307^{***}$    & $0.239^{***}$    & $0.322^{***}$    & $0.284^{***}$    & $0.252^{***}$    \\
Deep idiosyncratic & $0.391^{***}$    & $0.417^{***}$    & $0.386^{***}$    & $0.358^{***}$    & $0.388^{***}$    & $0.377^{***}$    & $0.423^{***}$    \\
PVAR(28)           & 0.755   & 0.992   & $0.606^{***}$   & $0.661^{***}$    & 0.890   & $0.785^{**}$    & 1.102   \\
                   \midrule
& \multicolumn{5}{c}{$h=14$} \\ \midrule
Deep pooled        &$0.304^{***}$    & $0.399^{***}$    & $0.340^{***}$    & $0.302^{***}$    & $0.353^{***}$    & $0.331^{***}$    & $0.295^{***}$    \\
Deep idiosyncratic & $0.459^{***}$    & $0.514^{***}$    & $0.422^{***}$    & $0.418^{***}$    & $0.428^{***}$    & $0.427^{***}$    & $0.449^{***}$    \\
PVAR(28)           & 0.773   & 1.035   & $0.592^{***}$    & $0.662^{***}$    & 0.876   & $0.806^{*}$    & 1.088   \\ 
\midrule
& \multicolumn{5}{c}{$h=21$} \\ \midrule
 Deep pooled        & $0.401^{***}$  & $0.461^{***}$  & $0.389^{***}$  & $0.365^{***}$  & $0.404^{***}$  & $0.391^{***}$  & $0.367^{***}$  \\
Deep idiosyncratic & $0.549^{***}$  & $0.569^{**}$  & $0.452^{***}$  & $0.495^{***}$  & $0.469^{***}$  & $0.496^{***}$  & $0.491^{***}$  \\
PVAR(28)           & $0.791$  & $1.036$  & $0.587^{***}$  & $0.658^{***}$  & $0.819^{*}$  & $0.833$  & $1.059$ 
     \\
     \bottomrule
		\end{tabular}
}
	\caption{Forecasting results, without the stringency index.  RMSE ratios, comparing the accuracy of each model against the deep time series model.  
		 Ratios $<1$ indicate superior  predictive ability of the respective model relative to the deep time-series model.  $*$, $**$, and
			$***$ denote  rejection of the null hypothesis of equality of forecast mean squared errors at the 10\%, 5\%, and 1\% levels of significance, respectively, using the modified \cite{Dieb95}  test with the \cite{harvey1997testing} adjustment.
}\label{dm1}
\end{table}

\begin{table}[H]
\centering
\resizebox{!}{!}{%
\begin{tabular}{@{}llllllll@{}}
\toprule
                      & Canada & France & Germany & Italy & Japan & UK    & US    \\ \midrule 
                  &  \multicolumn{5}{c}{$h=7$} \\ \midrule 
Deep pooled        & $0.290^{***}$   & $0.287^{***}$   & $0.295^{***}$    & $0.334^{***}$  & $0.270^{***}$  & $0.312^{***}$  & $0.321^{***}$  \\
Deep idiosyncratic & $0.469^{***}$   & $0.395^{***}$   & $0.372^{***}$    & $0.469^{***}$  & $0.311^{***}$  & $0.519^{***}$  & $0.540^{**}$  \\
PVAR(28)           & $0.780^{**}$   & 0.909  & $0.525^{***}$    & $0.638^{***}$  & $0.675^{***}$  & $0.728^{*}$  & $0.632^{**}$  \\
\midrule
& \multicolumn{5}{c}{$h=14$} \\ \midrule
Deep pooled        & $0.338^{***}$    & $0.395^{***}$    & $0.280^{***}$    & $0.379^{***}$   & $0.290^{***}$    & $0.342^{***}$           & $0.376^{***}$         \\
Deep idiosyncratic & $0.488^{***}$    & $0.488^{***}$    & $0.355^{***}$   & $0.494^{***}$    & $0.344^{***}$    & $0.561^{***}$           & $0.557^{***}$        \\
PVAR(28)           & $0.747^{**}$    & 0.968   & $0.507^{***}$    & $0.616^{***}$    & $0.648^{***}$    & $0.781^{*}$           & $0.683^{**}$         \\
\midrule
& \multicolumn{5}{c}{$h=21$} \\ \midrule
  Deep pooled        & $0.402^{***}$ & $0.462^{***}$ & $0.307^{***}$ & $0.442^{***}$ & $0.334^{***}$ & $0.400^{***}$ & $0.418^{***}$ \\
Deep idiosyncratic & $0.521^{***}$ & $0.543^{***}$ & $0.361^{***}$ & $0.531^{***}$ & $0.386^{***}$ & $0.596^{***}$ & $0.549^{***}$ \\
PVAR(28)           & $0.738^{***}$ & $0.990$ & $0.498^{***}$ & $0.609^{***}$ & $0.629^{***}$ & $0.791^{*}$ & $0.679^{**}$   \\
\bottomrule
		\end{tabular}
}
	\caption{Forecasting results with the aggregate Oxford stringency index. RMSE ratios, comparing the accuracy of each model against the deep time series model.  
		 Ratios $<1$ indicate superior  predictive ability of the respective model relative to the deep time-series model. For a description of the 4 forecasting models, see the notes to Table \ref{tbl7app}.  See further notes in Table \ref{dm1}
   % $*$, $**$, and
			% $***$ denote  rejection of the null hypothesis of equality of forecast mean squared errors at the 10\%, 5\%, and 1\% levels of significance, respectively, using the modified \cite{Dieb95}  test with the \cite{harvey1997testing} adjustment.
}\label{dm2}
\end{table}

\begin{table}[ht!]
\centering
\resizebox{!}{!}{%
\begin{tabular}{@{}llllllll@{}}
\toprule
& Canada & France & Germany & Italy & Japan & UK    & US    \\ \midrule 
                   & \multicolumn{5}{c}{$h=7$} \\ \midrule 
Deep pooled        & $0.235^{***}$   & $0.230^{***}$   & $0.245^{***}$    & $0.223^{***}$  & $0.257^{***}$  & $0.265^{***}$  & $0.199^{***}$  \\
Deep idiosyncratic & $0.374^{***}$   & $0.338^{***}$   & $0.329^{***}$    & $0.339^{***}$  & $0.319^{***}$  & $0.443^{***}$  & $0.372^{***}$  \\
PVAR(28)           & $0.700^{***}$   & $0.573^{**}$   & $0.562^{***}$    & $0.575^{***}$  & $0.673^{***}$  & $0.741^{*}$  & $0.757^{***}$  \\
\midrule
& \multicolumn{5}{c}{$h=14$} \\ \midrule
Deep pooled        & $0.299^{***}$    & $0.330^{***}$    & $0.276^{***}$    & $0.278^{***}$    & $0.277^{***}$    & $0.330^{***}$    & $0.256^{***}$    \\
Deep idiosyncratic & $0.392^{***}$    & $0.429^{***}$    & $0.378^{***}$    & $0.375^{***}$    & $0.342^{***}$    & $0.490^{***}$    & $0.398^{***}$   \\
PVAR(28)           & $0.684^{***}$    & $0.556^{**}$    & $0.579^{***}$    & $0.565^{***}$    & $0.641^{***}$    & $0.761^{*}$   & $0.763^{*}$    \\   
\midrule
& \multicolumn{5}{c}{$h=21$} \\ \midrule
 Deep pooled        & $0.363^{***}$  & $0.374^{***}$  & $0.294^{***}$  & $0.328^{***}$  & $0.314^{***}$  & $0.402^{***}$  & $0.323^{***}$  \\
Deep idiosyncratic & $0.427^{***}$  & $0.454^{**}$  & $0.401^{***}$  & $0.424^{***}$  & $0.385^{***}$  & $0.522^{***}$  & $0.420^{***}$  \\
PVAR(28)           & $0.661^{***}$  & $0.530^{**}$  & $0.566^{***}$ & $0.562^{***}$  & $0.639^{***}$  & $0.767^{}$  & $0.766^{***}$ \\   
\bottomrule
\end{tabular}
}
\caption{Forecasting results with the disaggregated Oxford stringency index. RMSE ratios, comparing the accuracy of each model against the deep time series model.  
		 Ratios $<1$ indicate superior  predictive ability of the respective model relative to the deep time-series model.   For a description of the 4 forecasting models, see the notes to Table \ref{tbl7app}.  See further notes in Table \ref{dm1}
}\label{dm3}
\end{table}

\subsection{Forecast evaluation in sub-samples}
In this section we present supplementary forecasting results as referenced in the main paper.  Specifically, we evaluate the forecasting performance of the proposed nonlinear panel estimator(s) relative to the two benchmark models, described in Section \ref{section4} of the main paper, over two distinct sub-periods within our overall out-of-sample window. The first sub-sample covers the period from February 6, 2021 to April 30, 2022, when COVID-19 was at its worst except in Japan, while the second is from May 1, 2022 through December 24, 2022.  Our aim is to examine whether policy mattered more in this earlier period, before immunity within each of the countries strengthened and COVID infection rates declined.

In Tables \ref{tbl7app4}--\ref{tbl7app3} we compare RMSE statistics across different models and countries  for these first and second sub-periods.  We do not include the stringency-based measures of policy and instead focus on predicting new COVID-19 cases using lags of new  cases and the other seven COVID-related measures.  We find across the forecasting horizons, $h\in \{7, 14, 21\} $ days, that the deep models yield significant forecasting gains over both the linear PVAR(28) model and the deep time-series neural network.  
 Similarly to the analysis in Section \ref{section4}, this shows the importance of both the panel dimension and of modeling nonlinearites when forecasting the
daily path of new COVID-19 cases across the G7 countries.   As anticipated, the RMSE values are smaller in the second sub-period, indicative of the lower COVID-19 transmission rates seen in Figure \ref{strinfig} from May 2022. 

In Tables \ref{tb4}--\ref{tb41} we present RMSE ratios, comparing the predictive ability of each  model with and without the aggregate Oxford stringency index over the two sub-samples. We see that policy as measured by the aggregate stringency index is more effective, with more RMSE ratios less than unity, in the latter sub-sample. But turning to the disaggregate stringency index, we see from Tables \ref{tb5}--\ref{tb51} that policy was then more effective even over the first sub-period. It is important to let the models choose how to weights the 9 components of the Oxford stringency index.

\begin{table}[!ht]
\centering
\resizebox{0.7\textwidth}{!}{%
\begin{tabular}{@{}lccccccc@{}}
\toprule
                   & Canada & France & Germany & Italy & Japan & UK    & US    \\ \midrule 
                   & \multicolumn{5}{c}{$h=7$} \\ \midrule 
Deep pooled        & 0.095 & 0.100 & 0.130 & 0.097 & 0.115 & 0.112 & 0.102 \\
Deep idiosyncratic & 0.145 & 0.134 & 0.158 & 0.130 & 0.138 & 0.171 & 0.158 \\
Deep time-series   & 0.325 & 0.362 & 0.465 & 0.301 & 0.448 & 0.349 & 0.313 \\
PVAR(28)           & 0.255 & 0.339 & 0.225 & 0.189 & 0.295 & 0.251 & 0.203\\ \midrule
& \multicolumn{5}{c}{$h=14$} \\ \midrule
Deep pooled        & 0.111 & 0.139 & 0.128 & 0.090 & 0.133 & 0.121 & 0.088 \\
Deep idiosyncratic & 0.147 & 0.159 & 0.141 & 0.129 & 0.157 & 0.140 & 0.115 \\
Deep time-series   & 0.366 & 0.347 & 0.397 & 0.298 & 0.362 & 0.362 & 0.305 \\
PVAR(28)           & 0.271 & 0.368 & 0.202 & 0.210 & 0.289 & 0.272 & 0.328 \\ \midrule
& \multicolumn{5}{c}{$h=21$} \\ \midrule
Deep pooled        & 0.144 & 0.156 & 0.148 & 0.115 & 0.159 & 0.142 & 0.118 \\
Deep idiosyncratic & 0.175 & 0.174 & 0.156 & 0.158 & 0.186 & 0.164 & 0.140 \\
Deep time-series   & 0.356 & 0.330 & 0.389 & 0.295 & 0.366 & 0.354 & 0.315 \\
PVAR(28)           & 0.270 & 0.353 & 0.185 & 0.206 & 0.276 & 0.276 & 0.332 \\
\bottomrule
\end{tabular}%
}
\caption{RMSE statistics for the 7, 14, and 21 day-ahead forecasts of new COVID-19 cases from the 4 models without policy-related variables over the sample February 6, 2021 to April 30, 2022.  }\label{tbl7app4}
 \end{table}

 \begin{table}[!ht]
\centering
\resizebox{0.70\textwidth}{!}{%
\begin{tabular}{@{}lccccccc@{}}
\toprule
                   & Canada & France & Germany & Italy & Japan & UK    & US    \\ \midrule 
                   & \multicolumn{5}{c}{$h=7$} \\ \midrule 
Deep pooled        & 0.030 & 0.060 & 0.084 & 0.080 & 0.124 & 0.033 & 0.039 \\
Deep idiosyncratic & 0.084 & 0.096 & 0.122 & 0.124 & 0.133 & 0.115 & 0.113 \\
Deep time-series   & 0.116 & 0.167 & 0.196 & 0.213 & 0.420 & 0.157 & 0.143 \\
PVAR(28)           & 0.087 & 0.100 & 0.166 & 0.143 & 0.298 & 0.127 & 0.064 \\ \midrule
& \multicolumn{5}{c}{$h=14$} \\ \midrule
Deep pooled        & 0.032 & 0.065 & 0.095 & 0.086 & 0.147 & 0.044 & 0.042 \\
Deep idiosyncratic & 0.120 & 0.141 & 0.153 & 0.109 & 0.185 & 0.109 & 0.114 \\
Deep time-series   & 0.103 & 0.172 & 0.220 & 0.283 & 0.440 & 0.153 & 0.119 \\
PVAR(28)           & 0.136 & 0.139 & 0.208 & 0.160 & 0.424 & 0.189 & 0.150 \\ \midrule
& \multicolumn{5}{c}{$h=21$} \\ \midrule
Deep pooled        & 0.032 & 0.065 & 0.094 & 0.086 & 0.153 & 0.046 & 0.041 \\
Deep idiosyncratic & 0.130 & 0.137 & 0.146 & 0.112 & 0.174 & 0.115 & 0.111 \\
Deep time-series   & 0.103 & 0.175 & 0.216 & 0.278 & 0.425 & 0.162 & 0.136 \\
PVAR(28)           & 0.133 & 0.135 & 0.220 & 0.158 & 0.381 & 0.193 & 0.151 \\
\bottomrule
\end{tabular}%
}
\caption{RMSE statistics for the 7, 14, and 21 day-ahead forecasts of new COVID-19 cases from the 4 models without policy-related variables over the sample May 1, 2022 to December 24, 2022.}\label{tbl7app3}
\end{table}

{\small
\begin{table}[!ht]
\centering
\resizebox{0.70\textwidth}{!}{%
\begin{tabular}{@{}llllllll@{}}
\toprule
		& Canada & France & Germany & Italy & Japan & UK    & US    \\ \midrule 
                 &  \multicolumn{5}{c}{$h=7$} \\ \midrule 
Deep pooled        & 1.096 & 0.934 & 1.092 & 1.466 & 1.013 & 1.064 & 1.387 \\
Deep idiosyncratic & 1.144 & 1.006 & 1.181 & 1.168 & 0.997 & 1.404 & 1.506 \\
Deep time-series   & 0.835 & 0.940 & 1.096 & 0.909 & 1.138 & 0.950 & 1.026 \\
PVAR(28)           & $0.903^{*}$ & $0.873^{**}$  & 0.973 & 0.842 & 0.930 & $0.911^{*}$  & $0.608^{**}$ \\ 
\midrule
& \multicolumn{5}{c}{$h=14$} \\ \midrule
Deep pooled        & 1.014 & $0.907^{*}$  & 0.900 & 1.248 & 0.931 & 0.947 & 1.261 \\
Deep idiosyncratic & 1.023 & 0.939 & 0.982 & 1.069 & 0.923 & 1.226 & 1.289 \\
Deep time-series   & 0.899 & 0.910 & 1.116 & 0.997 & 1.165 & 0.913 & 0.936 \\
PVAR(28)           & $0.908^{*}$  & $0.862^{***}$  & 0.980 & 0.843 & $0.913^{*}$  & 0.932 & $0.620^{**}$  \\ \midrule
& \multicolumn{5}{c}{$h=21$} \\ \midrule
Deep pooled        & 0.954 & $0.876^{*}$  & $0.850^{*}$  & 1.166 & 0.915 & 0.953 & 1.090 \\
Deep idiosyncratic & 0.952 & 0.884 & 0.884 & 0.945 & 0.886 & 1.151 & 1.078 \\
Deep time-series   & 0.944 & 0.865 & 1.117 & 0.993 & 1.141 & 0.938 & 0.934 \\
PVAR(28)           & $0.912^{*}$  & $0.844^{***}$  & 0.991 & 0.839 & $0.918^{*}$  & 0.932 & $0.624^{**}$ \\
\bottomrule	
\end{tabular}
}
	\caption{RMSE ratios, comparing the forecast accuracy of each respective model with and without the aggregate Oxford stringency index at 7, 14, and 21 days-ahead over the sample February 6, 2021 to April 30, 2022.  Ratios $<1$ indicate superior  predictive ability for the model with  the stringency index.  For a description of the 4 forecasting models, see the notes to Table \ref{tbl7app}.  $*$, $**$, and $***$ denote  rejection of the null hypothesis of equality of forecast mean squared errors with and without the aggregate Oxford stringency index at the 10\%, 5\%, and 1\% levels of significance, respectively, using the modified \cite{Dieb95}  test with the \cite{harvey1997testing} adjustment.}
 \label{tb4}
 \end{table}

 \begin{table}[!ht]
\centering
\resizebox{0.70\textwidth}{!}{%
\begin{tabular}{@{}llllllll@{}}
\toprule
		& Canada & France & Germany & Italy & Japan & UK    & US    \\ \midrule 
                 &  \multicolumn{5}{c}{$h=7$} \\ \midrule 
Deep pooled        & 0.916 & 0.900 & 0.853 & 0.911 & 0.741 & 0.764 & 0.896 \\
Deep idiosyncratic & $0.683^{*}$ & $0.677^{*}$ & $0.793^{*}$ & 1.120 & $0.661^{*}$ & 1.052 & 0.987 \\
Deep time-series   & 1.103 & 1.005 & 0.920 & 0.793 & 0.902 & 0.948 & 1.193 \\
PVAR(28)           & $0.621^{***}$ & $0.769^{*}$ & $0.826^{*}$ & $0.879^{*}$ & $0.645^{**}$ & 0.722 & $0.454^{***}$ \\ 
\midrule
& \multicolumn{5}{c}{$h=14$} \\ \midrule
Deep pooled        & 0.954 & 0.906 & 0.880 & 0.935 & 0.742 & 1.029 & 0.883 \\
Deep idiosyncratic & 0.795 & $0.672^{**}$ & $0.793^{*}$ & 1.132 & 0.717 & 1.160 & 0.981 \\
Deep time-series   & 1.138 & 0.954 & 0.877 & 0.736 & 0.873 & 0.985 & 1.212 \\
PVAR(28)           & $0.640^{***}$ & $0.762^{*}$ & $0.838^{*}$ & 0.895 & $0.630^{**}$ & 0.714 & $0.426^{***}$ \\ \midrule
& \multicolumn{5}{c}{$h=21$} \\ \midrule
Deep pooled        & 0.926 & 0.917 & 0.928 & 0.947 & 0.727 & 1.022 & 0.894 \\
Deep idiosyncratic & 0.721 & $0.685^{*}$ & 0.853 & 1.133 & 0.772 & 1.021 & 1.005 \\
Deep time-series   & 1.093 & 0.953 & 0.945 & 0.762 & 0.871 & 0.922 & 1.062 \\
PVAR(28)           & $0.669^{***}$ & $0.775^{*}$ & $0.842^{*}$ & $0.899^{**}$ & $0.659^{*}$ & 0.709 & $0.428^{***}$ \\
\bottomrule	
\end{tabular}
}
	\caption{RMSE ratios, comparing the forecast accuracy of each respective model with and without the aggregate Oxford stringency index at 7, 14, and 21 days-ahead over the sample May 1, 2022 to December
24, 2022.  Ratios $<1$ indicate superior  predictive ability for the model with  the stringency index.  For a description of the 4 forecasting models, see the notes to Table \ref{tbl7app}.  $*$, $**$, and $***$ denote  rejection of the null hypothesis of equality of forecast mean squared errors with and without the aggregate Oxford stringency index at the 10\%, 5\%, and 1\% levels of significance, respectively, using the modified \cite{Dieb95}  test with the \cite{harvey1997testing} adjustment.
 % $*$, $**$, and $***$ denote  rejection of the null hypothesis of equality of forecast mean squared errors with and without the aggregate Oxford stringency index at the 10\%, 5\%, and 1\% levels of significance, respectively, using the modified \cite{Dieb95}  test with the \cite{harvey1997testing} adjustment.
 }
 \label{tb41}
\end{table}
}

\begin{table}[!ht]
\centering
\resizebox{0.70\textwidth}{!}{%
\begin{tabular}{@{}clllllll@{}}
\toprule
		& Canada & France & Germany & Italy & Japan & UK    & US    \\ \midrule 
                 &  \multicolumn{5}{c}{$h=7$} \\ \midrule 
Deep pooled        & 0.934 & $0.847^{**}$ & $0.880^{*}$ & 1.027 & 0.963 & $0.852^{*}$ & 0.867 \\
Deep idiosyncratic & $0.821^{**}$ & 0.914 & 1.029 & 0.843 & 1.010 & 1.017 & 0.892 \\
Deep time-series   & 0.860 & 1.053 & 1.041 & 0.928 & 1.087 & 0.864 & 1.074 \\
PVAR(28)           & 0.870 & $0.591^{***}$ & 1.044 & 0.741 & 1.023 & 0.887 & 0.764 \\ 
\midrule
& \multicolumn{5}{c}{$h=14$} \\ \midrule
Deep pooled        & 0.915 & $0.895^{*}$ & $0.889^{*}$ & 0.932 & 0.901 & $0.894^{*}$ & 0.929 \\
Deep idiosyncratic & $0.775^{**}$ & 0.914 & 1.039 & $0.792^{*}$ & 0.942 & 0.957 & 0.864 \\
Deep time-series   & 0.886 & 1.079 & 1.034 & 0.970 & 1.108 & 0.838 & 1.040 \\
PVAR(28)           & 0.855 & $0.549^{***}$ & 1.128 & 0.720 & 1.032 & 0.881 & 0.764 \\ \midrule
& \multicolumn{5}{c}{$h=21$} \\ \midrule
Deep pooled        & 0.897 & $0.886^{*}$ & $0.859^{**}$ & $0.878^{**}$ & $0.880^{*}$ & 0.939 & $0.898^{**}$ \\
Deep idiosyncratic & $0.777^{**}$ & 0.893 & 1.066 & $0.769^{**}$ & 0.885 & 0.927 & $0.834^{*}$ \\
Deep time-series   & 0.951 & 1.096 & 1.027 & 0.969 & 1.095 & 0.860 & 1.009 \\
PVAR(28)           & 0.859 & $0.521^{***}$ & 1.218 & 0.716 & 1.060 & 0.870 & 0.759 \\
\bottomrule	

	\end{tabular}
}
	\caption{RMSE ratios, comparing the forecast accuracy of each respective model with and without the disaggregate Oxford stringency index at 7, 14, and 21 days-ahead over the sample February 6, 2021 to April 30, 2022.  Ratios $<1$ indicate superior  predictive ability for the model with  the stringency index.  For a description of the 4 forecasting models, see the notes to Table \ref{tbl7app}.  $*$, $**$, and $***$ denote  rejection of the null hypothesis of equality of forecast mean squared errors with and without the aggregate Oxford stringency index at the 10\%, 5\%, and 1\% levels of significance, respectively, using the modified \cite{Dieb95}  test with the \cite{harvey1997testing} adjustment.}\label{tb5}
\end{table}

 \begin{table}[!ht]
\centering
\resizebox{0.70\textwidth}{!}{%
\begin{tabular}{@{}llllllll@{}}
\toprule
		& Canada & France & Germany & Italy & Japan & UK    & US    \\ \midrule 
                 &  \multicolumn{5}{c}{$h=7$} \\ \midrule 
Deep pooled        & 1.020 & 0.771 & 0.769 & $0.600^{*}$ & $0.743^{*}$ & $0.734^{*}$ & $0.719^{**}$ \\
Deep idiosyncratic & 0.950 & 0.763 & $0.717^{**}$ & 0.893 & 0.733 & 1.152 & 1.027 \\
Deep time-series   & 1.712 & 1.150 & 1.272 & 0.854 & 1.039 & 1.192 & 1.081 \\
PVAR(28)           & $0.558^{***}$ & 0.914 & 0.852 & 0.950 & $0.547^{**}$ & 0.652 & $0.373^{***}$ \\ 
\midrule
& \multicolumn{5}{c}{$h=14$} \\ \midrule
Deep pooled        & 1.027 & 0.870 & 0.795 & $0.627^{*}$ & 0.796 & 0.797 & $0.717^{**}$ \\
Deep idiosyncratic & 0.852 & 0.867 & 0.807 & 0.893 & 0.782 & 1.195 & 1.029 \\
Deep time-series   & 1.698 & 1.084 & 1.243 & 0.799 & 1.081 & 1.309 & 1.097 \\
PVAR(28)           & $0.568^{***}$ & 0.879 & 0.878 & 0.956 & $0.535^{**}$ & 0.661 & $0.356^{***}$ \\ \midrule
& \multicolumn{5}{c}{$h=21$} \\ \midrule
Deep pooled        & 1.007 & 0.872 & 0.781 & $0.624^{*}$ & 0.766 & $0.764^{*}$ & $0.716^{**}$ \\
Deep idiosyncratic & 0.759 & 0.795 & 0.819 & 0.848 & 0.903 & 1.042 & 0.943 \\
Deep time-series   & 1.660 & 1.047 & 1.541 & 0.809 & 1.068 & 1.227 & 1.015 \\
PVAR(28)           & $0.578^{***}$ & 0.890 & 0.869 & 0.965 & $0.579^{**}$ & 0.688 & $0.398^{***}$ \\
\bottomrule	

	\end{tabular}
}
	\caption{RMSE ratios, comparing the forecast accuracy of each respective model with and without the disaggregate Oxford stringency index at 7, 14, and 21 days-ahead over the sample May 1, 2022 to December
24, 2022.  Ratios $<1$ indicate superior  predictive ability for the model with  the stringency index.  For a description of the 4 forecasting models, see the notes to Table \ref{tbl7app}.  $*$, $**$, and $***$ denote  rejection of the null hypothesis of equality of forecast mean squared errors with and without the aggregate Oxford stringency index at the 10\%, 5\%, and 1\% levels of significance, respectively, using the modified \cite{Dieb95}  test with the \cite{harvey1997testing} adjustment.
 }\label{tb51}
\end{table}

\clearpage
\subsection{Temporal instabilities in forecast performance: the fluctuation test}\label{GRTest}

\begin{figure}[!ht]		
 \centering
	\includegraphics[width=\textwidth]{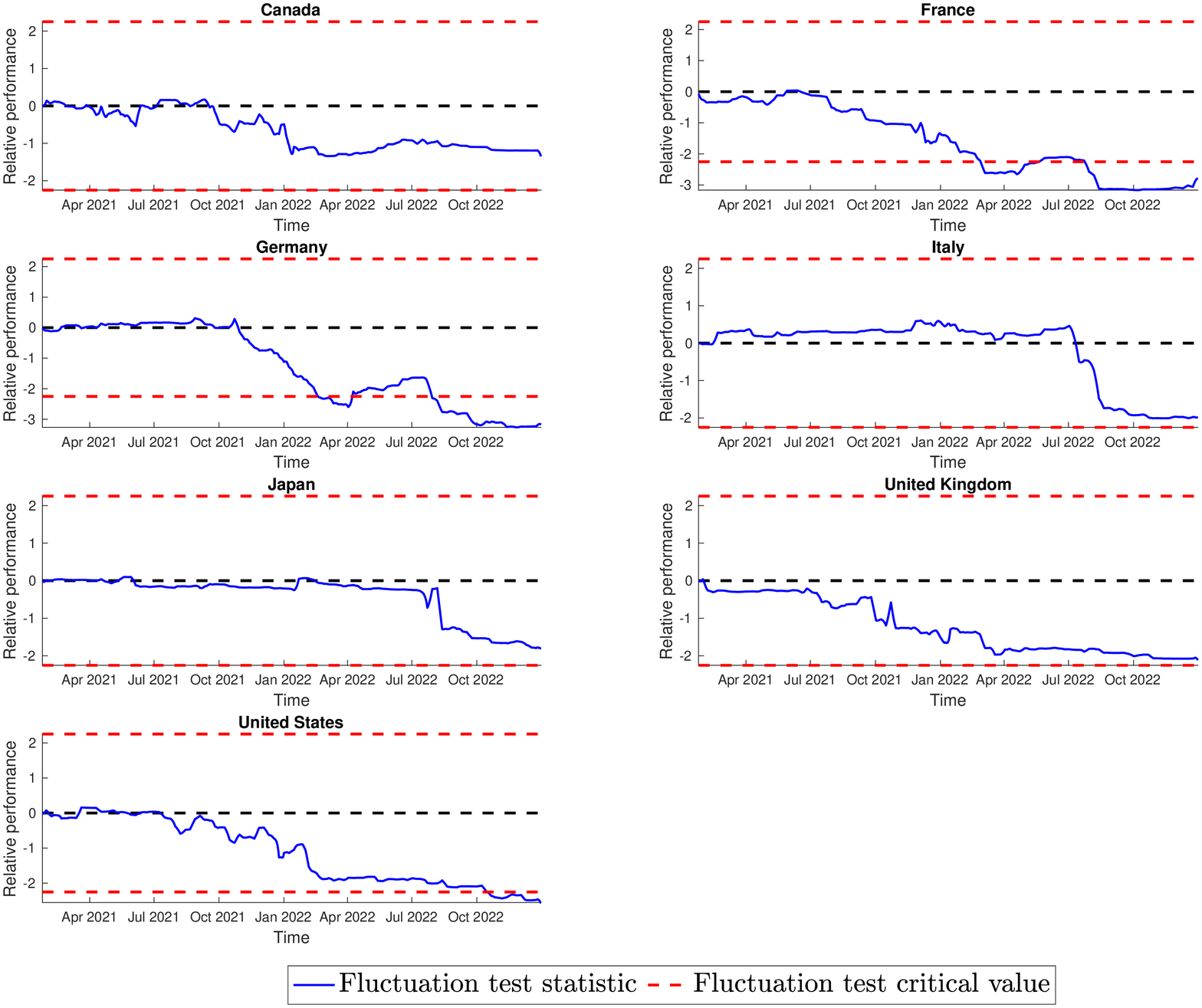}
	\caption{Giacomini and Rossi's (2010) fluctuation test, obtained as the (standardized) difference between the MSE of the deep pooled model with and without the disaggregated Oxford stringency index at $h=7$. Negative values of the fluctuation statistic imply that the model with the disaggregate stringency index is better. Critical values are at the $10\%$ level of significance.}\label{fig8app}		
\end{figure}

To compare the predictive performance of competing models in unstable environments, \cite{giacomini2010forecast} propose the fluctuation test. It utilizes the test statistic of \cite{Dieb95} computed  over rolling out-of-sample windows of size $m$. Given the evidence in Table \ref{tb5} that the disaggregate Oxford stringency index improves the forecasts from the deep pooled model -- on average over the period February 6, 2021 through December 24, 2022 -- Figure \ref{fig8app} uses the fluctuation test to test the null hypothesis that the local RMSE  equals zero at each point in time. When the test statistic (the solid blue line) crosses the critical values (the dashed red line) equal forecast performance is rejected. 

\newpage
\clearpage
\subsection{Empirical evidence using penalized models}

In this section we present forecasting results adding an $\ell_1$ penalty on the weights of each corresponding network estimator; see Section \ref{section3} of the main paper.  We follow the same forecasting design  as in Section \ref{fdesign} of the main paper.  

We start by presenting the results from the penalized estimation. 
In Table \ref{tbl9app} we report the RMSE ratio of each model with the aggregate Oxford stringency index as considered in Table \ref{tb4a}, i.e., deep pooled, deep idiosyncratic, and deep time series versus the  deep pooled  lasso, deep idiosyncratic  lasso, and deep time series  lasso. Ratios less than one indicate superior predictive ability for the model without the lasso penalization.   
% Highlighted are the entries corresponding to the  $\textit{RMSE}_i$  of the Deep Pooled lasso, Deep idiosyncratic  lasso   and Deep time-series with lasso.   
In Table \ref{tbl10app} we report the same metrics as in  Table \ref{tbl9app}, but consider the disaggregated stringency index as considered in Table \ref{tb5a} in the main paper.  

In both Tables \ref{tbl9app}--\ref{tbl10app} the evidence is compelling.     We find that the  more heavily parameterized models -- deep pooled, deep idiosyncratic, and deep time-series -- forecast better without penalization.
On the face of it, this seems quite surprising, given that in many other contexts penalized models have been found to forecast well. But this finding can be understood in relation to the recent statistical  literature on so called \textit{double descent}; see \cite{hastie2022surprises} and \cite{kelly2022virtue}.  We discuss this issue further in Remark \ref{remark_double_descen} in Section \ref{section3} of the main paper.

\begin{table}[!ht]
\centering
\rowcolors{2}{gray!25}{white}
\resizebox{0.8\linewidth}{!}{%
\begin{tabular}{@{}lccccccc@{}}
\toprule
	% \centering
	% \rowcolors{2}{gray!25}{white}
	% \begin{tabular}{@{}lccccccc@{}}
		& Canada & France & Germany & Italy & Japan & UK    & US    \\ \midrule
		Deep pooled              & 0.809  & 0.740  & 0.670   & 0.782 & 0.620 & 0.736 & 0.772 \\
		Deep pooled lasso        & 0.097  & 0.120  & 0.173   & 0.117 & 0.191 & 0.126 & 0.111 \\
		Deep idiosyncratic       & 0.864  & 0.723  & 0.741   & 0.776 & 0.604 & 0.975 & 0.811 \\
		Deep idiosyncratic lasso & 0.147  & 0.169  & 0.197   & 0.165 & 0.226 & 0.158 & 0.177 \\
		Deep time-series         & 0.854  & 1.047  & 0.989   & 0.884 & 0.923 & 1.057 & 1.016 \\
		Deep time-series lasso   & 0.318  & 0.294  & 0.397   & 0.310 & 0.475 & 0.281 & 0.262 \\
		PVAR(28)                 & 0.212  & 0.280  & 0.206   & 0.175 & 0.296 & 0.216 & 0.168\\ \bottomrule
	\end{tabular}
 }
	\caption{RMSE ratios, comparing the forecast accuracy of each respective model with the aggregate Oxford stringency index 7 days-ahead when estimated with and without penalization.   Entries $<1$ indicate superior  predictive ability of the model with  no penalty.  }
	\label{tbl9app}
\end{table}

\begin{table}[!ht]
\centering
\rowcolors{2}{gray!25}{white}
\resizebox{0.8\linewidth}{!}{%
\begin{tabular}{@{}lccccccc@{}}
\toprule
	% \centering
	% \rowcolors{2}{gray!25}{white}
	% \begin{tabular}{@{}lccccccc@{}}
	% 	\toprule
		& Canada & France & Germany & Italy & Japan & UK    & US    \\ \midrule
		Deep pooled              & 0.728  & 0.600  & 0.581   & 0.653 & 0.541 & 0.664 & 0.628 \\
		Deep pooled lasso        & 0.094  & 0.133  & 0.164   & 0.096 & 0.213 & 0.113 & 0.087 \\
		Deep idiosyncratic       & 0.746  & 0.733  & 0.591   & 0.604 & 0.604 & 0.851 & 0.718 \\
		Deep idiosyncratic lasso & 0.146  & 0.160  & 0.218   & 0.159 & 0.236 & 0.147 & 0.143 \\
		Deep time-series         & 1.048  & 1.081  & 0.984   & 0.989 & 1.023 & 0.978 & 0.812 \\
		Deep time-series lasso   & 0.278  & 0.320  & 0.397   & 0.286 & 0.438 & 0.288 & 0.339 \\
		PVAR(28)                 & 0.203  & 0.198  & 0.219   & 0.162 & 0.301 & 0.209 & 0.209 \\ \bottomrule
	\end{tabular}
 }
	\caption{RMSE ratios, comparing the forecast accuracy of each respective model with the disaggregate Oxford stringency index 7 days-ahead when estimated with and without penalization.   Entries $<1$ indicate superior  predictive ability of the model with  no penalty. }
	\label{tbl10app}
\end{table}

\clearpage
\section{The effectiveness of policy: Disaggregated partial derivatives}

This section presents plots of the partial derivatives, (\ref{partial}), for those disaggregated stringency measures from the Oxford index not shown in the main paper.

\begin{figure}[!ht]		
\centering
	\includegraphics[width=\linewidth]{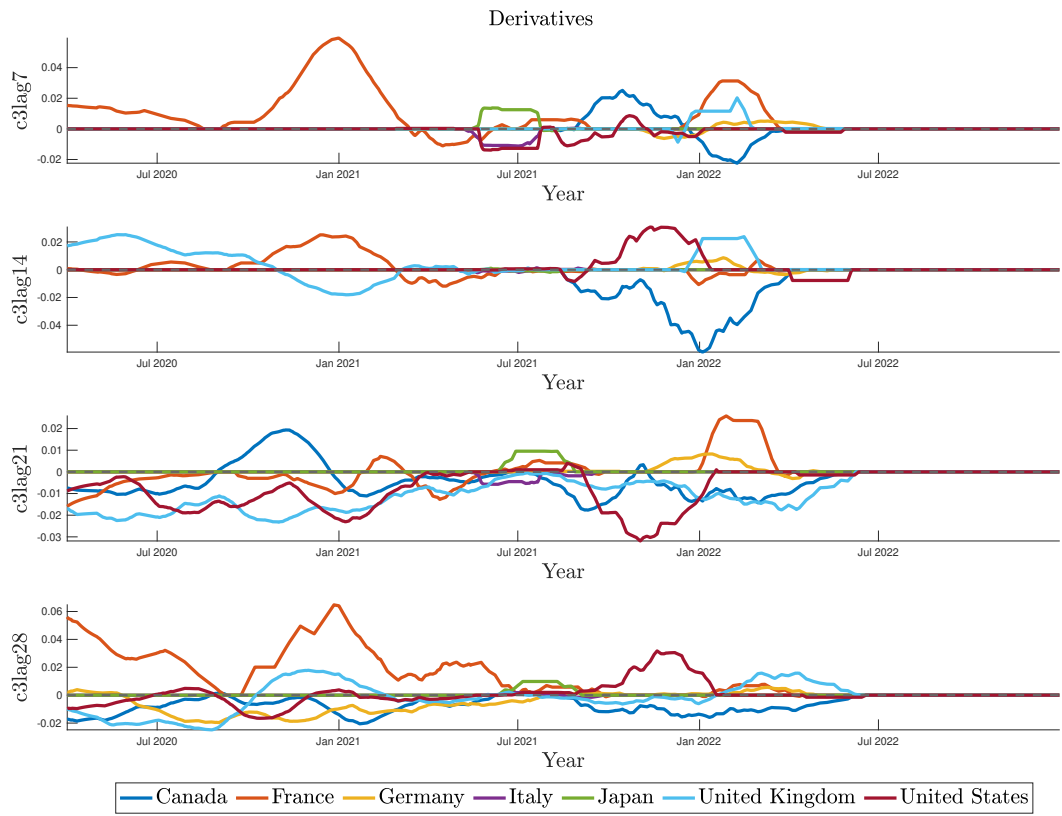}
	\caption{Partial derivatives: The effects of cancelling public events
on new COVID-19 cases 7, 14, 21, and 28 days after the policy change}	
\end{figure}	
% \clearpage\newpage

\begin{figure}[!ht]		
\centering
	\includegraphics[width=0.95\textwidth]{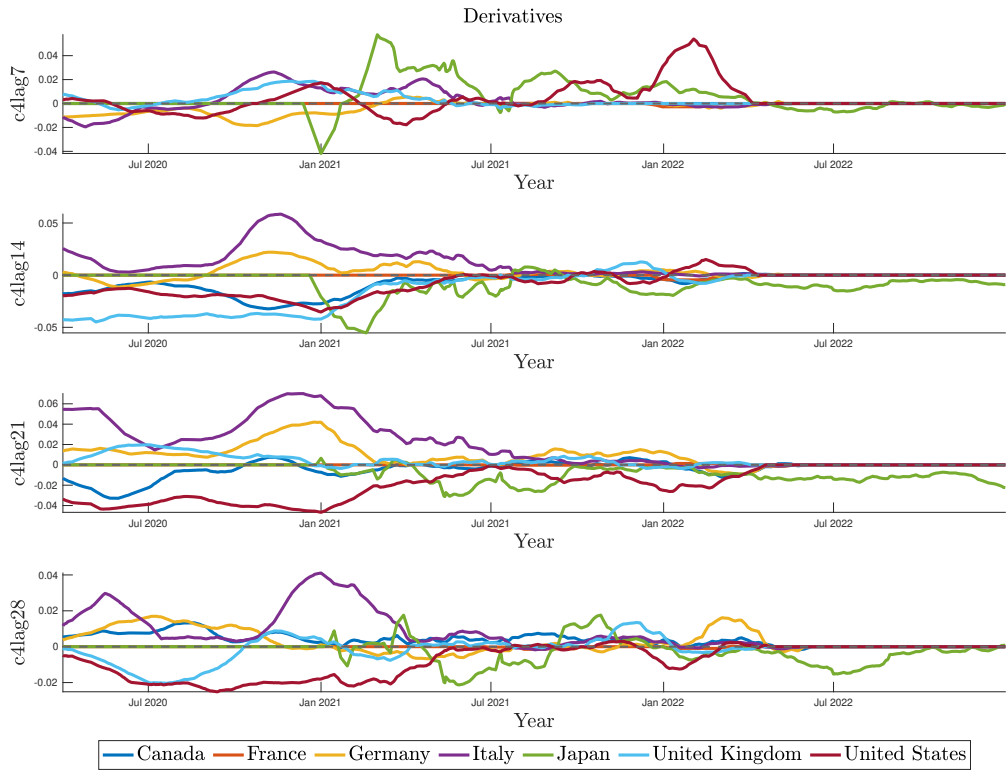}
	\caption{Partial derivatives: The effects of imposing limits on gatherings
on new COVID-19 cases 7, 14, 21, and 28 days after the policy change}	
\end{figure}	
% \clearpage\newpage

\begin{figure}[!ht]		
\centering
	\includegraphics[width=0.95\textwidth]{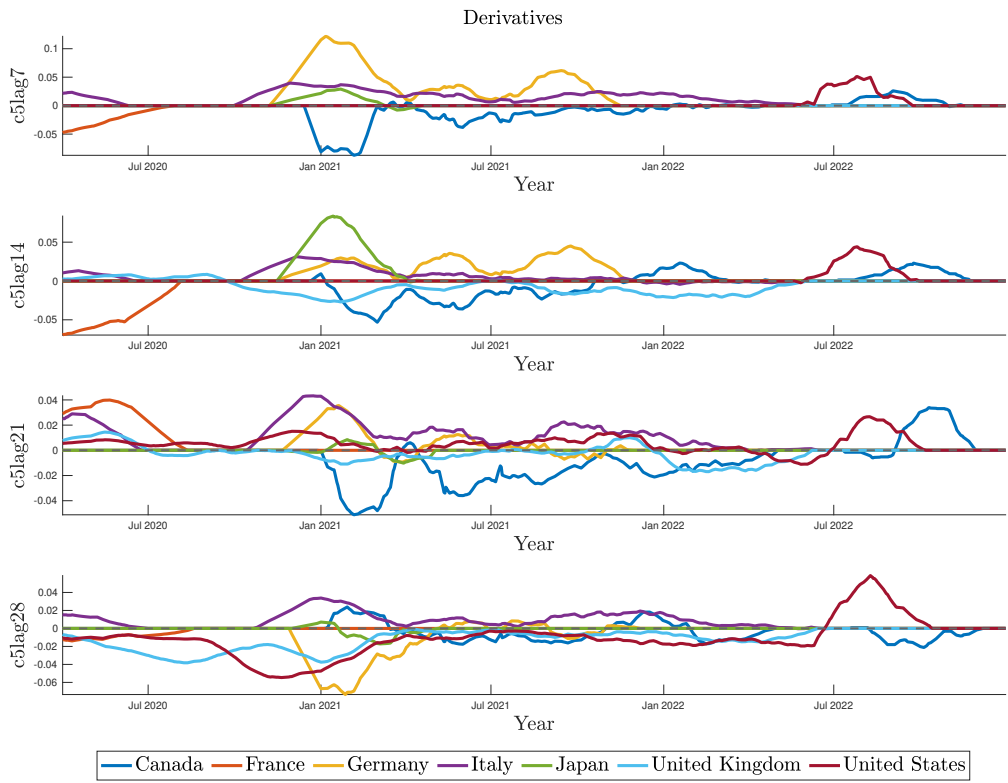}
	\caption{Partial derivatives: The effects of closing public transport
on new COVID-19 cases 7, 14, 21, and 28 days after the policy change}	
\end{figure}	
%\clearpage\newpage

\begin{figure}[!ht]		
\centering
	\includegraphics[width=0.95\linewidth]{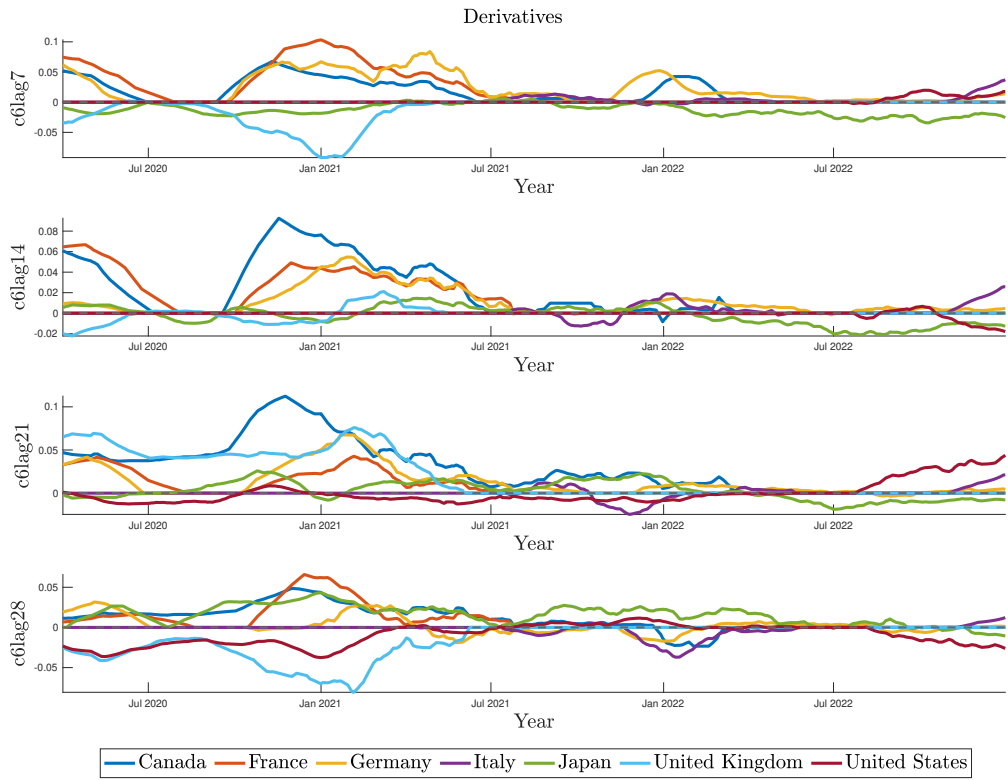}
	\caption{Partial derivatives: The effects of orders to ``shelter-in-place'' and other stay-at-home orders
on new COVID-19 cases 7, 14, 21, and 28 days after the policy change }	
\end{figure}	
% \clearpage\newpage

\begin{figure}[!ht]		
\centering
	\includegraphics[width=0.95\linewidth]{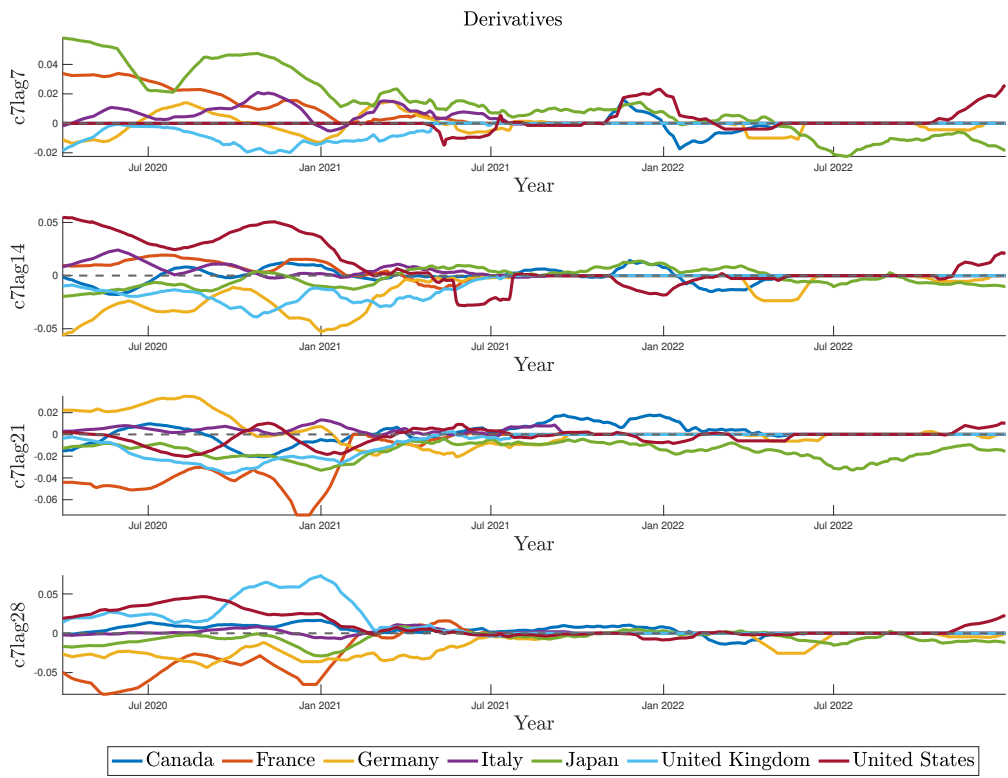}
	\caption{Partial derivatives: The effects of restrictions on internal movement between cities/regions
on new COVID-19 cases 7, 14, 21, and 28 days after the policy change}	
\end{figure}	
% \clearpage\newpage

\begin{figure}[!ht]		
\centering
	\includegraphics[width=0.9\linewidth]{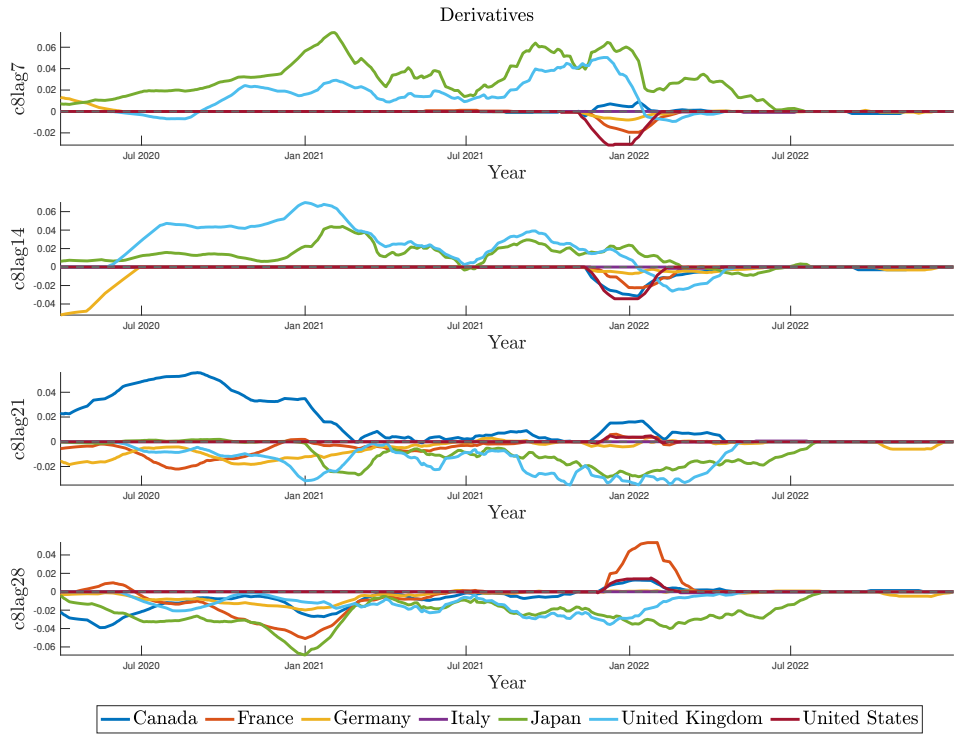}
	\caption{Partial derivatives: The effects of restrictions on international travel
on new COVID-19 cases 7, 14, 21, and 28 days after the policy change. Note: this records policy for foreign travellers, not citizens.}	
\end{figure}	
%\clearpage\newpage

\begin{figure}[!ht]		
\centering
	\includegraphics[width=0.9\linewidth]{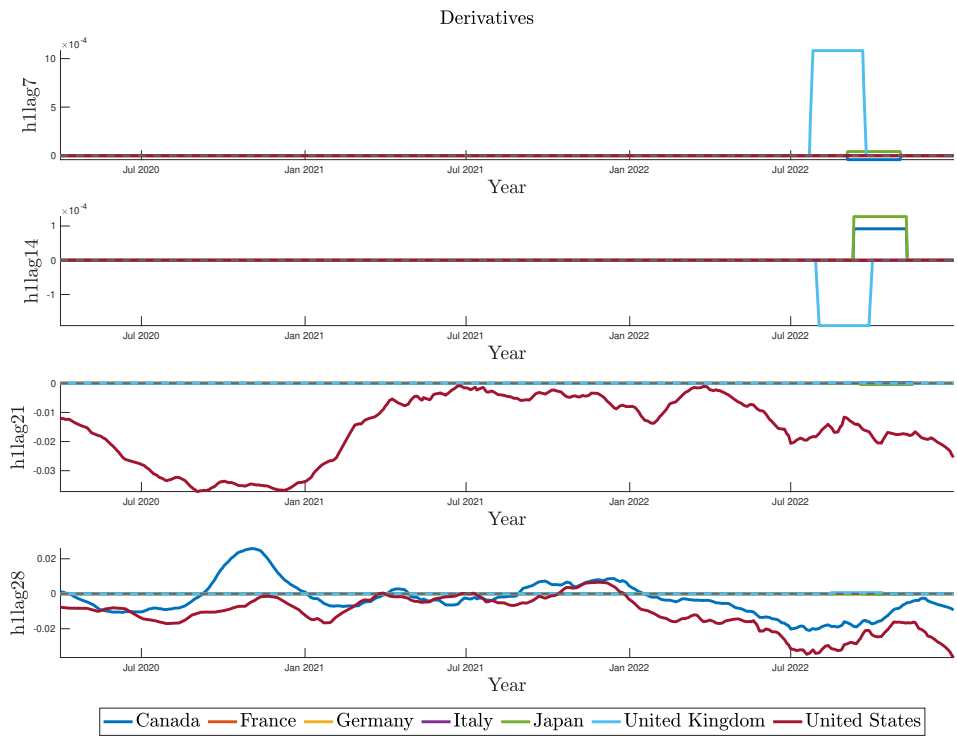}
	\caption{Partial derivatives: The effects of public information campaigns
on new COVID-19 cases 7, 14, 21, and 28 days after the policy change.
Note: no differentiated policies reported in this indicator.}	
\end{figure}	
%\clearpage\newpage

\end{document}